\documentstyle[12pt,a41,epsfig]{article}

\newcommand{\ds}{\displaystyle}
\newcommand{\ab}{\overline{\alpha}_s}

\newcommand{\gsim}{\raisebox{-0.07cm}{$\,\stackrel{>}{{\scriptstyle
 \sim}}\, $} }
\newcommand{\lsim}{\raisebox{-0.07cm}{$\,\stackrel{<}{{\scriptstyle
 \sim}}\, $} }

\newcommand\MSbar{$\overline{\mbox{\rm MS}}\,$}

\newcommand\GeV{\,\mbox{GeV}}

\newcommand\beq{\begin{equation}}
\newcommand\eeq{\end{equation}}
\newcommand\bea{\begin{eqnarray}}
\newcommand\eea{\end{eqnarray}}
\newcommand\ra{\rightarrow}

\newcommand\FV{\,\mbox{\boldmath $F$}}
\newcommand\CV{\,\mbox{\boldmath $C$}}
\newcommand\qV{\mbox{\boldmath $q$}}
\newcommand\eV{\mbox{\boldmath $e$}}

\newcommand\GGV{\,\mbox{\boldmath $\gamma$}}

\newcommand\PV{\mbox{\boldmath $P$}}
\newcommand\RV{\mbox{\boldmath $R$}}
\newcommand\KV{\mbox{\boldmath $K$}}
\newcommand\kV{\mbox{\boldmath $k$}}

\newcommand\UV{\,\mbox{\boldmath $U$}}
\newcommand\IV{\,\mbox{\boldmath $I$}}
\newcommand\LV{\,\mbox{\boldmath $L$}}

\newcommand\GAM{\,\mbox{\boldmath $\gamma$}}

\begin{document}
\setlength{\baselineskip}{0.515cm}
\sloppy
\thispagestyle{empty}
\begin{flushleft}
DESY 96--096 \hfill
{\tt hep-ph/9712546}\\
WUE-ITP-96-39 \hfill December 1997
\end{flushleft}

\setcounter{page}{0}

\mbox{}
\vspace*{\fill}
\begin{center}
{\LARGE\bf The Evolution of Unpolarized Singlet}

\vspace{2mm}
{\LARGE\bf Structure Functions at Small \mbox{\boldmath $x$}} \\

\vspace{5em}
\large
Johannes Bl\"umlein$^a$ and Andreas Vogt$^b$
\\
\vspace{5em}
\normalsize
{\it $^a$DESY--Zeuthen}\\
{\it Platanenallee 6, D--15735 Zeuthen, Germany}\\

\vspace{5mm}
{\it $^b$Institut f\"ur Theoretische Physik, Universit\"at W\"urzburg}\\
{\it Am Hubland, D--97074 W\"urzburg, Germany}\\
\vspace*{\fill}
\end{center}
\begin{abstract}
\noindent
A systematic study is performed of the impact of the various resummed
small-$x$ contributions to the anomalous dimensions and coefficient
functions on the evolution of unpolarized structure functions in
deep-inelastic scattering. The proton structure functions $F_2^p$ and
$F_L^p$ as well as the photon structure function $F_2^{\gamma}$ are
considered together with the corresponding parton densities. The
general analytic solution of the evolution equations in Mellin-$N$
space is derived, and different approximate solutions are compared.
Potential effects of less singular small-$x$ terms in the anomalous
dimension and coefficient functions are discussed.
\end{abstract}

\vspace{1mm}
\noindent
\begin{center}
PACS numbers~:~13.60.Hb, 12.38.Cy, 12.38bx
\end{center}

\vspace*{\fill}
\newpage
\section{Introduction}
\label{sect1}
\setcounter{equation}{0}

\noindent
One of the central questions in the theory of deep-inelastic scattering
(DIS) structure functions is that of their behavior at small values of
Bjorken-$x$. The HERA experiments~\cite{HEXP} have performed detailed
measurements of the structure function $F_2(x,Q^2)$ down to values
of $x \simeq 10^{-5}$ and have presented first results on $F_L(x,Q^2)$
in the range $x \gsim 10^{-4}$~\cite{FLEXP}. $F_2$ rises even at small
values of $Q^2 \sim 2 \GeV^2$ approximately as $x^{-0.2}$. At low
scales $Q^2$ the behavior of structure functions cannot be dealt with
by means of perturbative QCD due to the size of the strong coupling
constant $\alpha_s(Q^2)$. For large virtualities, on the other hand, a
perturbative description of the scaling violations is possible if a
factorization can be achieved between the non-perturbative input
distributions and the evolution kernels which can be calculated
perturbatively.

Throughout the present paper we will consider deep-inelastic scattering
in the range where it is dominated by the light-cone singularities in
the Bjorken limit, $Q^2, s \rightarrow \infty, \, x = Q^2/(sy) = {\sf
const.}$, with $Q^2$ the 4-momentum transfer squared, $s$ the
center-of-mass energy, $y = 2 P\cdot q/s$.
The ultraviolet singularities of the operators emerging in the
light-cone expansion~\cite{LCE} are associated to renormalization
group equations (RGE) which describe the evolution of the structure
functions. Under these conditions the evolution kernels are given by
the anomalous dimensions of the light-cone operators. For the
leading-twist contributions considered in the following, the
expectation values of the operators are related to the parton
distribution functions. This picture holds, in principle, down to the
region of small values of $x$. The anomalous dimensions and coefficient
functions, however, may receive large low-$x$ contributions of the
type~\cite{LIPAT}
$$
 \alpha_s^l \left ( \frac{\alpha_s}{N - 1} \right )^k
 \:\:\: \leftrightarrow \:\:\:
 \frac{1}{x} \,  \alpha_{s}^{l+k} \, \frac{\ln^{k-1}(1/x)}{(k-1)!}\:\: ,
$$
where $N$ denotes the index of the Mellin--transformation
$$
 {\cal M}\left [f(x)\right](N) = \int_0^1 \! dx \, x^N f(x)~.
$$
The $x$-range in which this representation is applicable can only be
determined by explicit calculations~\cite{AHM} and depends on both the
behavior of the partonic input densities at a starting scale $Q^2_0$
and on the structure of the anomalous dimensions and coefficient
functions. Note that the present picture does not necessarily yield a
description of the structure functions as well in the Regge limit $s
\ra \infty, Q^2~=~{\sf const.}$, since both limits lead in general to
different results. This can be seen~\cite{RL} performing these limits
in the Jost--Lehmann--Dyson representation~\cite{JLD} of the structure
functions.

In the Bjorken limit the ultraviolet and collinear divergences emerging
in the calculation of the higher-order corrections can be dealt with
applying the corresponding RGE-operators, which imply the evolution of
the parton densities and the running of the strong coupling constant.
The resummed small-$x$ corrections in leading~\cite{LIPAT} and
next-to-leading
order~[8--10]
can thus be accounted for in
a natural way. Since the impact of the resulting all-order anomalous
dimensions on the behavior of the DIS structure functions at small $x$
does as well depend on the non--perturbative input parton densities at
an initial scale $Q_0^2$, the perturbative resummation effects can
only be studied via the evolution over some range in $Q^2$. This
evolution probes also the anomalous dimensions and coefficient
functions at medium and large values of $x$ due to the  Mellin
convolution between the evolution kernels and the parton densities,
cf.~ref.~\cite{JBDUR}. Hence the small-$x$ dominance of the leading
terms over less singular contributions as $x \ra 0 $ in the anomalous
dimensions and coefficient functions does not necessarily imply the
same effect for the observables, such as the structure functions.

In the present paper a systematic study is performed on the impact
of the different resummed small-$x$ contributions to the anomalous
dimensions and coefficient functions on the evolution of the singlet
contributions to the nucleon structure functions $F_2(x,Q^2)$ and
$F_L(x,Q^2)$.
A brief summary of some of our numerical results on the effects of the
resummed gluon anomalous dimension on the nucleon structure functions
has already appeared in~\cite{BV97}. We extend the analysis to the
photon structure function $F_2^{\gamma}(x,Q^2)$, for which we derive
corresponding results in the DIS scheme.

The paper is organized as follows. The general framework for the
evolution of parton densities of the nucleon and the photon is outlined
in Section~2. In Section~3 the presently known small-$x$ resummed
anomalous dimensions and coefficient
functions~[4,8--10]
are summarized and the numerical
coefficients for their expansions in $\alpha_s$ are compiled for the
subsequent analysis. The issue of less singular terms is discussed in
Section~4 guided by the known 2--loop results. In Section~5 different
methods are derived for the solution of the evolution equations in the
presence of all--order resummations for the small-$x$ contributions.
The numerical implications of the small-$x$ resummations on the
evolution of the parton densities and the structure functions $F_2^p$,
$F_L^p$ and $F_2^\gamma$ are worked out in detail in Section~6.
Section~7 contains our conclusions.

\section{The evolution equations}
\label{sect2}
\setcounter{equation}{0}

\noindent
The twist-2 contributions to the structure functions in inclusive
deep-inelastic scattering can be described in terms of the QCD-improved
parton model. Their scaling violations are governed by renormalization
group equations which can be formulated to all orders in the strong
coupling constant. In this section we briefly recall this general
framework, which allows for a consistent introduction of the small-$x$
resummations into the structure function evolution, and its specific
application to hadronic and photonic parton distributions.

\subsection{The renormalization group equations}
\label{sect21}

\noindent
Among the singularities emerging in the calculation of QCD radiative
corrections, only the ultraviolet divergences and the initial-state
mass singularities require a special treatment in inclusive
deep-inelastic scattering$\, $%
\footnote{The infrared divergences cancel between the virtual and
 Bremsstrahlung contributions according to the Bloch--Nordsieck theorem
 \cite{BLNO}. The final-state mass singularities vanish due to the
 Kinoshita--Lee--Nauenberg theorem~\cite{KLN}, as all degenerate final
 states are summed over.}.
The former are eliminated by the renormalization of the strong coupling
constant. The remaining mass singularities, originating in collinear
emissions of massless partons off massless partons, are removed by mass
factorization, i.e., these contributions are absorbed into the bare
parton densities. For this procedure the structure functions
$F_i(x,Q^2)$ are first written as
\beq
\label{eqFi1}
 F_i(x,Q^2) = \hat{F}_{i,k} \bigg(x,\alpha_s(R^2), \frac{Q^2}{\mu^2},
 \frac{R^2}{\mu^2}, \varepsilon \bigg) \otimes \hat{f}_k(x) \: .
\eeq
Here $\alpha_s(R^2)$ denotes the strong coupling constant, renormalized
at the scale $R$ in some renormalization scheme. $\mu$ is an arbitrary
mass scale. $\hat{f}_k$ represents the bare momentum distribution
of the parton species $k$, and $\otimes $ stands for the Mellin
convolution in the first variable,
\beq
\label{conv}
 A(x,\mu_a^2) \otimes B(x,\mu_b^2) = \int_0^1 \! dx_1 \int_0^1 \!
 dx_2 \, \delta(x - x_1 x_2) \, A(x_1,\mu_a^2) B(x_2,\mu_b^2) \: .
\eeq
Finally $\varepsilon$ marks the initial-state mass singularities
entering the bare partonic structure functions~$\hat{F}_{i,k}$. These
functions are then separated into the coefficients functions $C_{i,j}$
and the transition functions $\Gamma_{jk}$, which contain the
$1/\varepsilon$ pole terms, according to
\beq
\label{eqFi2}
 \hat{F}_{i,k} \bigg( x, \alpha_s(R^2), \frac{Q^2}{\mu^2}, \frac{R^2}
 {\mu^2}, \varepsilon \bigg) = C_{i,j} \bigg( x, \alpha_s(R^2),
 \frac{Q^2}{M^2}, \frac{R^2}{M^2} \bigg) \otimes \Gamma_{jk} \bigg(
 x, \alpha_s(R^2), \frac{M^2}{\mu^2}, \frac{M^2}{R^2}, \varepsilon
 \bigg) \: .
\eeq
The additional parameter $M$ is the factorization scale. This separation
is not unique beyond the leading order (LO) of the perturbative
expansion, hence $\Gamma_{kj}$ and $C_{i,k}$ are also factorization
scheme dependent.  Combining Eqs.~(\ref{eqFi1}) and (\ref{eqFi2}), the
structure functions $F_i(x,Q^2)$ finally read
\beq
\label{eqFi3}
 F_i(x,Q^2)  =  C_{i,j} \bigg( x,\alpha_s(R^2), \frac{Q^2}{M^2},
 \frac{R^2}{M^2} \bigg ) \otimes f_j \bigg( x,\alpha_s(R^2),
 \frac{M^2}{\mu^2}, \frac{M^2}{R^2} \bigg)
\eeq
in terms of the renormalized parton densities $f_j$ given by
\beq
 \label{eqFi4}
 f_j \bigg( x, \alpha_s(R^2), \frac{M^2}{\mu^2}, \frac{M^2}{R^2} \bigg)
 =  \Gamma_{jk} \bigg( x, \alpha_s(R^2), \frac{M^2}{\mu^2},
 \frac{M^2}{R^2}, \varepsilon \bigg) \otimes \hat{f}_k (x) \: .
\eeq

Two arbitrary scales, $R$ and $M$, are thus introduced by the
renormalization and mass factorization procedures. These scales are
not physical. Hence observables, such as the structure functions
$F_i(x, Q^2)$ in Eq.~(\ref{eqFi3}), do not depend on them. Since the
ultraviolet and mass singularities are not related, the conditions
\bea
\label{eqmur}
  R^2 \,\frac{d}{d R^2}\, F_i(x,Q^2) &=& 0 \: , \\
  M^2 \frac{d}{d M^2} F_i(x,Q^2) &=& 0
\label{eqmuf}
\eea
hold separately and imply two independent renormalization group
equations. The first of these equations leads to the scale dependence
of the running coupling constant,
\beq
 \label{eval}
 \frac{da_s}{d\ln R^2} = \beta(a_s) \equiv
 - \sum_{l=0}^{\infty} a^{l+2}_s \beta_l \: ,
\eeq
where the abbreviation $a_s = \alpha_s(R^2)/(4\pi)$ has been introduced
for convenience. At next-to-leading order (NLO) only the first two,
scheme independent terms \cite{beta0,beta1} of $\beta (a_s)$ are kept,
\bea
 \beta_0 &=& \frac{11}{3}\, C_A   - \frac{4}{3}\,  T_F \: ,\nonumber \\
 \beta_1 &=& \frac{34}{3}\, C_A^2 - \frac{20}{3}\, C_A T_F
      - 4\, C_F T_F \: .
\label{beta1}
\eea
The QCD color factors are $ C_F = (N_c^2-1)/(2 N_c) \equiv 4/3 $,
$ C_A = N_c \equiv 3 $, $T_R = 1/2$, and $ T_F = N_f T_R $, with $N_f$
denoting the number of light quark flavors. To this approximation,
the solution of Eq.~(\ref{eval}) can be expressed in terms of the QCD
scale parameter $\Lambda_{N_f}$ by
\beq
 \label{alps}
 \frac{1}{\beta_0 a_s} - \frac{\beta_1}{\beta_0^2} \ln \left(
 \frac{1}{\beta_0 a_s} + \frac{\beta_1}{\beta_0^2} \right)  =
 \ln \left( \frac{R^2}{\Lambda_{N_f}^2} \right) \: .
\eeq
In turns out, moreover, that higher coefficients $\beta_{l \geq 2}$ --
$\beta_2$ and $\beta_3$ have been calculated in the \MSbar\  scheme
\cite{beta2,beta3} -- are not required in connection with the
presently available small-$x$ resummations, see Sect.~5.3. We will
therefore employ the relation (\ref{alps}) for all our numerical
calculations in Sect.~6.

\subsection{Hadronic and photonic parton densities}
\label{sect22}

\noindent
The second renormalization group equation (\ref{eqmuf}) leads to the
scale evolution of the renormalized parton densities $f_j(x,M^2)$.
Considering first the hadronic case, the relevant parton species are
the quarks and antiquarks, $q_j$ and $\overline{q}_j$, and the gluon
$g$. It is convenient to introduce flavor non-singlet combinations of
the quark densities,
\bea
 q_{j}^{\pm} &=& q_j \pm \overline{q}_j - \frac{1} {N_f}
  \sum_{r=1}^{N_f} \, [q_r \pm \overline{q}_r] \: ,  \\
 q^{\rm val} &=& \sum_{r=1}^{N_f}\, [q_r - \overline{q}_r] \: ,
\label{NSpar}
\eea
and the singlet quark$\,$/$\,$gluon vector

\beq
\label{Spar}
 \qV = \left( \begin{array}{c} \!\Sigma\! \\ \! g\! \end{array}\right)
 \:\: ,\:\: \Sigma \equiv \sum_{r=1}^{N_f} \, [q_{r}+\bar{q}_{r}] \: .
\eeq
This decomposition decouples the $ 2 N_f\! +\! 1 $ evolution equations
as far as possible by symmetry considerations alone. For simplicity,
we will choose the renormalization and factorization scales as $R^2 =
M^2 = Q^2$ from now on$\, $%
\footnote{See refs.~\cite{MRSCH,RH2} for studies of the uncertainties
 in NLO analyses due to the variation of $R$ and $M$.}.
The hadronic evolution equations can then be written as
\bea
 \frac{\partial q_j^{\pm}(x,Q^2)}{\partial \ln  Q^2}
 &=& P^{\pm}(x, a_s) \otimes q_j^{\pm}(x,Q^2) \: , \nonumber \\
 \:\frac{\partial \qV (x,Q^2)}{\partial \ln  Q^2}\:
 &=& \:\PV (x, a_s) \,\otimes\, \qV (x,Q^2) \: .
\label{evol1}
\eea
The evolution of $q^{\rm val}$ is identical to that of $q^-$ up to NLO.
As our present analysis is confined to that approximation in the
non-singlet sector$\, $%
\footnote{The generating relations for the resummation of the anomalous
 dimensions~\cite{BVplb1} of the $\pm$ non-singlet combinations were
 derived in ref.~\cite{KL}. Here the leading small-$x$ terms are of
 $O[(\alpha_s \ln^2 x)^n]$.
 The effect of these terms has turned out to be on the
 1\% level down to very small values of $x$~\cite{BVplb1,BRV}. As the
 non-singlet contributions are furthermore suppressed compared to the
 singlet ones at low $x$, these resummations are not included in the 
 present treatment.},
the corresponding equation has been suppressed in Eqs.~(\ref{evol1}).
In general, the splitting functions $P^{\pm}$, $\PV$ are given by the
infinite series
\bea
 P^{\pm}(x,a_s)
   &\! =\! &  \sum_{l=0}^{\infty} a_{s}^{l+1} P_l^{\pm}(x)
   \: , \nonumber \\
 \:\PV (x,a_s)\:
   &\!\equiv \!& \left( \begin{array}{cc}
               \! P_{qq}(x,a_s) \! & \! P_{qg}(x,a_s) \! \\
               \! P_{gq}(x,a_s) \! & \! P_{gg}(x,a_s) \!
                        \end{array} \right)
     \, = \,  \sum_{l=0}^{\infty} a_s^{l+1} \PV_{l}(x) \: .
\label{Ps}
\eea
The expansion coefficients $ P^{-}_{l}(x)$ and $\PV_{l}^{\rm unpol}(x)$
are, in sensible factorization schemes, subject to the sum rules
\bea
\label{fcons}
  & & \int_0^1 \! dx \, P_{l}^-(x) = 0 \: , \\
\label{conserv}
  & & \int_0^1 \! dx \, x \sum_j P_{jk,l}^{\rm unpol.}(x) = 0 \: ,
\eea
which are due to fermion number and energy-momentum conservation,
respectively. By now all unpolarized and polarized entries in Eqs.\
(\ref{Ps}) are completely known at NLO, $l=1$. The full expressions for
their $x$-dependences can be found in
refs.~[24--28].
Beyond this order a series of integer Mellin moments of $P_2^+(x)$ and
$\PV_2(x)$ has been calculated so far \cite{LOOP3}.

We now turn to the parton densities of the real photon. The photon is
a genuine elementary particle, unlike the hadrons. Hence it can
directly take part in hard scattering processes, in addition to its
quark and gluon distributions arising from quantum fluctuations,
$q^{\gamma}(x,Q^2)$ and $g^{\gamma}(x,Q^2)$. Denoting the corresponding
photon distribution in the photon by $\Gamma^{\,\gamma}(x,Q^2)$, the
evolution equations for these parton densities are generally given
by~\cite{DWi}
\bea
  \frac{\partial q_i^{\,\gamma}}{\partial \ln Q^{2}} &\! =\! &
  a_{\rm em} \overline{P}_{q_{i}\gamma} \otimes \Gamma^{\,\gamma}
  \, + \, a_s \,\bigg\{ 2 \sum_{k=1}^{N_f} \overline{P}_{q_{i}q_{k}}
          \otimes q_{k}^{\,\gamma}
  \, + \,\overline{P}_{q_{i}g} \otimes g^{\gamma} \bigg\} \: ,
\nonumber \\
\label{evol11}
  \frac{\partial g^{\gamma}}{\partial \ln Q^{2}} &\! =\! &
  a_{\rm em} \overline{P}_{g\gamma} \:\otimes \Gamma^{\,\gamma}
  \, + \, a_s \,\bigg\{ 2 \sum_{k=1}^{N_f} \overline{P}_{gq_{k}} \:
          \otimes q_{k}^{\,\gamma}
  \, + \,\overline{P}_{gg} \:\otimes g^{\gamma} \bigg\} \: ,  \\
  \frac{\partial\,\Gamma^{\,\gamma}}{\partial \ln Q^{2}} &\! =\! &
  a_{\rm em} \overline{P}_{\gamma\gamma} \:\otimes \Gamma^{\,\gamma}
  \, + a_{\rm em} \bigg\{ 2 \sum_{k=1}^{N_f} \overline{P}_{\gamma
          q_{k}} \:\otimes q_{k}^{\,\gamma}
  \, + \,\overline{P}_{\gamma g} \:\otimes g^{\gamma} \bigg\} \: .
\nonumber
\eea
Here $a_{\rm em} \equiv \alpha / (4\pi)$ with the electromagnetic
coupling constant $\alpha \simeq 1/137$. The antiquark distributions do
not occur separately in Eqs.~(\ref{evol11}), as $\bar{q}_i ^{\,\gamma}
(x,Q^2) = q_i^{\,\gamma}(x,Q^2)$ due to charge conjugation invariance.
The generalized splitting functions read
\beq
\label{pgen}
  \overline{P}_{ij}(x,a_{\rm em},a_s) = \sum_{l,m=0} a_{\rm em}^m
  a_s^l \,\overline{P}_{ij}^{\, (m,l)}(x) \: ,
\eeq
with $\overline{P}_{q_{i}q_{k}}$ being the average of the quark-quark
and antiquark-quark splitting functions.

Usually calculations involving the photon's parton structure are
restricted to the first order in $\alpha \ll 1$. In this approximation
all $m\neq 0$ terms in Eq.~(\ref{pgen}) can be neglected, since
$q_i^{\,\gamma}$ and $g^{\gamma}$ are already of order $\alpha$. This
reduces the functions $\overline {P}_{ij}$ to the usual QCD quantities
${P}_{ij}(x,\alpha_s)$, and ${P}_{\gamma q_i}$ and ${P}_{\gamma g}$
drop out completely. Moreover one has ${P}_{\gamma \gamma} \propto
\delta (1-x)$ to all orders in $\alpha_s$, as real photon radiation
from photons starts at order $\alpha^2$ only. Thus the last line of
Eq.~(\ref {evol11}) can be integrated immediately, at leading order
(LO, $l=0$), for example, resulting in
\beq
 \Gamma_{\rm LO}^{\,\gamma}(x,Q^2) = \delta (1-x) \Big[ 1 - 4a_s \Big(
 \sum_{k=1}^{N_f} e_{q_k}^2 \,\ln \frac{Q^2}{Q_0^2} + \mbox{const.}
 \Big) \Big] \: ,
\eeq
where $e_{q_k}$ represents the quark charges, and $Q_0^2$ is some
reference scale for the evolution. Only the $O(1)$ part of $\Gamma^
{\,\gamma}$ affects the quark and gluon densities at order $\alpha $,
as well as any observable involving hadronic final states like
$F_2^{\,\gamma}$. Therefore, after decomposing into the singlet and
non-singlet parts as before, one obtains the inhomogeneous evolution
equations
\bea
 \frac{\partial q_j^{\gamma +}(x,Q^2)}{\partial \ln  Q^2}
 &=& k_j^{+}(x,a_s) + P^+(x, a_s) \otimes q_j^{\gamma +}(x,Q^2)
 \: , \nonumber \\
 \:\frac{\partial \qV^{\gamma} (x,Q^2)}{\partial \ln  Q^2}\:
 &=& \: \,\kV (x,a_s)\, + \:\PV (x, a_s) \:\otimes\, \qV^{\gamma}
 (x,Q^2) \: .
\label{evolx}
\eea
Here we have switched to the conventional notation for the photon-parton
splitting functions
\bea
 k^{+}(x,\alpha_s) &=& \sum_{l=0}^{\infty} a_{\rm em} a_s^l \,
   k^{+}_l(x) \: , \nonumber \\
 \kV (x,a_s) \, &\equiv & \left( \begin{array}{c}
             \! P_{q\gamma }(x,a_s) \! \\
             \! P_{g\gamma }(x,a_s) \! \end{array} \right)
  =  \sum_{l=0}^{\infty} a_{\rm em} a_s^l \, \kV _l(x) \: .
\eea
These splitting functions are presently also known up to NLO ($l=1$),
see
refs.~[30--32].

\subsection{Factorization scheme transformations}
\label{sect23}

\noindent
As stated above, the separation between the coefficient functions and
the splitting functions is not unique beyond the leading order. In this
subsection, we derive the general factorization-scheme invariants and
specify the schemes for our subsequent numerical calculation. We study
the photonic case, as the hadronic problem forms a subset hereof. One
way to carry out this task is to introduce a fictitious second,
`gluonic' structure function, where the gluon density enters at order
$a_s^0$, but the quarks only at order $a_s^1$, opposite to the situation
with the real electromagnetic current, see refs.~\cite{FP,GR83}. This
can be achieved, e.g., by adding a color-neutral scalar-gluon coupling
$\phi F^{\mu\nu} F_{\mu\nu}$ to the QCD Lagrangian. The general singlet
structure function $F_2$ then reads
\beq
  \FV_2 =
  \left( \begin{array}{c} \!F_{2,\gamma}\! \\ \! F_{2,\phi}\!
  \end{array} \right) = \: <\! e^2\! > [\CV(x,a_s) \otimes \qV(x,a_s)
  + \CV_{\gamma}(x,a_s)]
\eeq
with
\bea
 \CV(x,a_s)\, & = & \left( \begin{array}{cc}
     \! C_{\gamma q}(x,a_s)  & \! C_{\gamma g}(x,a_s) \! \\
     \! C_{\phi q}(x,a_s)    & \! C_{\phi g}(x,a_s)   \!
     \end{array} \right)
   \, = \, \sum_{l=0}^{\infty} a_s^l \CV_l(x) \: , \nonumber \\
 \CV_{\gamma}(x,a_s) & = & \left( \begin{array}{c}
     \! C_{\gamma \gamma}(x,a_s) \! \\
     \! C_{\phi \gamma}(x,a_s) \! \end{array} \right) \, = \,
   \sum_{l=0}^{\infty} a_{\rm em} a_s^l \CV_{\gamma ,l+1}(x) \: .
\eea
Here $<\! e^2 \! >$  is the average squared charge of the light quark
flavors,  and $\CV_0 $ the unit matrix times $\delta(1-x)$.
Using the scale dependence (\ref{evolx}) of the partons and Eq.\
(\ref{eval}) for the running coupling, the scaling violations of $F_2$
can be written as
\bea
 \frac{d\FV_2}{d \ln Q^2} &=& <\! e^2 \! > [\CV \otimes \kV + \beta
 \CV_{\gamma}^{\prime} - (\CV \otimes \PV \otimes \CV^{-1} + \beta
 \CV^{\prime} \otimes \CV^{-1}) \otimes \CV_{\gamma}] \nonumber \\
 & & \mbox{} + (\CV \otimes \PV \otimes \CV^{-1} + \beta \CV^{\prime}
 \otimes \CV^{-1}) \otimes \FV_2 \: ,
\label{strf}
\eea
where the prime denotes the derivative with respect to $a_s$. Both
$d\FV_2 /d \ln Q^2$ and $\FV_2$ represent observables, hence the
following combinations of splitting functions and coefficient functions
are factorization scheme invariant:
\bea
 \IV_{\rm hom.}\, &=& \CV \otimes \PV \otimes \CV^{-1} + \beta
   \CV^{\prime} \otimes \CV^{-1} \: ,\\
\label{sinv1}
 \IV_{\rm inhom} &=& \CV \otimes \kV  + \beta \CV_{\gamma} -
   \IV_{\rm hom.} \otimes \CV_{\gamma}  \: .
\label{sinv}
\eea
Putting back the perturbative expansions of all quantities involved,
these invariants are given order-by-order in $a_s$ \cite{FP,GR83,GRVph1}
by
\bea
\label{eqi1}
 \IV_{\rm hom.}\, &=& \, a_s \PV_0 \, + \, a_s^2 \{ \PV_1 + \CV_1
  \otimes \PV_0 - \PV_0 \otimes \CV_1 - \beta_0 \CV_1 \} + \ldots\: ,\\
 \IV_{\rm inhom} &=& a_{\rm em} \kV_0 + \, a_{\rm em} a_s \{ \kV_1 +
  \CV_1 \kV_0 - \PV_0 \CV_{\gamma , 1} \} + \ldots \:\: .
\label{eqi2}
\eea
The generalization to higher orders is straightforward if
cumbersome. From
these relations the changes $\Delta \PV $ and $\Delta \kV$ of the
splitting functions induced by a modification of the coefficients
$\Delta \CV$ and $\Delta \CV_{\gamma}$ can be easily determined. Recall,
however, that a particular choice for the physical upper-row
(electromagnetic) quantities $C_{\gamma q}$, $C_{\gamma g}$ and
$C_{\gamma\gamma}$ does not fully fix the transformation, as well-known
for the hadronic DIS factorization scheme \cite{DIS}. In fact, we will
use the DIS scheme for our subsequent numerical calculations in both
the hadronic and the photonic cases. Starting from the usual \MSbar\
scheme of fixed-order
calculations~[36--38,29],
the transformation reads
\renewcommand{\arraystretch}{1.3}
\beq
  \Delta \CV_1 = \left( \begin{array}{rr}
       \! -C_{\gamma q, 1}^{{\overline{\rm MS}}}(x)  &
          -C_{\gamma g, 1}^{{\overline{\rm MS}}}(x) \! \\
       \!  C_{\gamma g, 1}^{{\overline{\rm MS}}}(x)  &
           C_{\gamma g, 1}^{{\overline{\rm MS}}}(x) \!
  \end{array} \right) \:\: , \:\:
   \Delta \CV_{\gamma ,1} = \left( \begin{array}{c}
       \! -C_{\gamma\gamma , 1}^{{\overline{\rm MS}}}(x) \! \\
       \!          0                     \!
  \end{array} \right) \: .
\label{Ctrf}
\eeq
\renewcommand{\arraystretch}{1}
\noindent
The lower-row choice in the hadronic part is the conventional
continuation  of the sum-rule constraint to all $N$, that one of the
photonic part is taken over from the DIS$_{\gamma}$ scheme of
ref.~\cite{GRVph1}, see also Sect.~3.3.
This concludes the general all-order framework, and we can now turn
to the resummed anomalous dimensions and coefficient functions.

\section{Small-{$x$} resummations for the anomalous
 dimensions and coefficient functions}
\label{sect3}
\setcounter{equation}{0}

\noindent
In this section we briefly summarize relations for the resummed singlet
anomalous dimensions in Mellin-$N$ space which are used in the numerical
analysis below. The anomalous dimension matrix is related to the
corresponding singlet splitting functions by
\beq
 \GGV (N,a_s) = -2 \PV(N,a_s) \,\equiv\,
 -2 \int_0^1 \! dx \, x^{N} \PV (x,\alpha_s)~.
\eeq
The general form of the small-$x$ resummed, unpolarized anomalous
dimension matrix reads
\beq
 \GGV_{\rm res}(N,\alpha_s)  =
    \sum_{k=0}^{\infty} \left( \frac{\ab}{N} \right)^{k+1}
    \sum_{m=0} \ab^{\, m} \GAM_k^{(m)}~,
\label{gares}
\eeq
with $\ab = C_A \alpha_s(Q^2)/\pi$. In the discussion below we include
the LO~\cite{SpF1} and NLO~\cite{SpF4} anomalous dimensions $\GGV_0$
and $\GGV_1$ completely, and account for the small-$x$ resummed series
in the leading (L$x$, $m=0$) and next-to-leading (NL$x$, $m=1$)
small-$x$ approximations
\beq
\label{gres}
 \GGV (N,\alpha_s)  = a_s \GGV_0 (N) +  a_s^2 \GGV_1(N)
 +  \sum_{k=2}^{\infty} \left( \frac{\ab}{N} \right)^{k+1}
 \left [ \GAM_k^{(0)} + N \GAM_k^{(1)} + O(N^2) \right ]~.
\eeq

\subsection{The leading series}
\label{sect31}

\noindent
The all-order resummation of the L$x$ series was performed in refs.\
\cite{LIPAT}:
\beq
\label{GAML}
\GGV_L(N, \alpha_s) =   -2
\sum_{k=0}^{\infty} \left (\begin{array}{cc}
 g_{k,qq}^{(0)}
& g_{k,qg}^{(0)}  \vspace{1mm}
\\  g_{k,gq}^{(0)} & g_{k,gg}^{(0)} \end{array}
\right ) \left(\frac{\ab} {N} \right )^{k+1}
= -2  \left( \begin{array}{cc} \! 0 \!       & \! 0 \! \\
                               \! C_F/C_A \! & \! 1 \! \end{array}
\right) \gamma _L (N, \alpha_s )
\eeq
with $\gamma_L(N,\alpha_s)$ being the solution of
\beq
\label{LIP}
\rho \equiv
 \frac{N}{\overline{\alpha}_s} = 2 \psi(1) - \psi(\gamma_L)
 - \psi(1 - \gamma_L) \equiv  \chi_0(\gamma_L)~.
\eeq
$\psi(z)$ denotes the logarithmic derivative of Euler's
$\Gamma$-function. $\gamma_L $ is a multi-valued function for complex
values of $N$. The perturbative branch of the solution is selected by
the requirement~\cite{BCM}
\beq
 \gamma_L(N,\alpha_s) \rightarrow \frac{\overline{\alpha}_s}{N}
 ~~~~{\rm for}~~|N| \rightarrow \infty~.
\label{phys}
\eeq
For small values of $\gamma_L(N,\alpha_s)$ the asymptotic representation
\begin{equation}
\gamma_L = \frac{\overline{\alpha}_s}{N} \left \{ 1 +
2 \sum_{l = 1}^{\infty} \zeta(2 l + 1) \gamma_L^{2l +1} \right\}
\end{equation}
holds, from which the coefficients $g_{k,gg}^{(0)}$ in Eq.~(\ref{GAML})
can be determined iteratively. Here $\zeta(n)$ denotes the Riemann
$\zeta$-function. Note that only $\zeta$-functions of odd integers
contribute, which is expected for physical quantities in four
dimensions, cf.\ ref.~\cite{KREI}. Analytic expressions for $g_{k,gg}^
{(0)}$ were given up to $k=14$ in ref.~\cite{CFM}. Later both analytic
representations and the numerical values of these coefficients were
determined to large values of $k$ by various authors (cf., e.g.,
refs.~\cite{BF1,JBKT}). In fact, the series (\ref{GAML}) can be used
as representation of $\gamma_L(N, \alpha_s)$ along typical integration
contours (cf.\ Fig.~4 in Sect.~5) for the inverse Mellin transform, as
shown in ref.~\cite{JBKT}: 20 terms (or less) in Eq.~(\ref{GAML}) are
sufficient to obtain an accuracy of better than 0.01\%.

For the later numerical analysis, i.e., to perform the Mellin inversion
of the solution of the evolution equations, it is necessary to locate
the singularities of $\gamma_L(N, \alpha_s)$ in the complex $N$ plane.
The singularities of the resummed leading singular part of the
anomalous dimension $\gamma_L$ can be determined by differentiating
eq.~(\ref{LIP}) with respect to $\rho$,
\begin{equation}
 \frac{d \gamma_L}{d \rho} \left [ - \psi'(\gamma_L)
 + \psi'(1 - \gamma_L) \right ] = 1~.
\end{equation}
The condition
\begin{equation}
\left [\frac{d \gamma_L}{d \rho} \right ]^{-1} =
\psi'(\gamma_L) - \frac{\pi^2}{2} \frac{1}{\sin^2(\pi \gamma_L)} = 0
\label{cond}
\end{equation}
yields the value of the resummed anomalous dimension $\gamma_L$ at the
branch points. The corresponding values of $\rho$ are then determined
by Eq.~(\ref{LIP}). For the perturbative branch one obtains
\renewcommand{\arraystretch}{1.5}
\begin{equation}
\begin{array}{ll}
\gamma_1 \:\, = \ds \frac{1}{2}  &~~~~ \rho_1 \:\, = 4 \ln 2~, \\
\gamma_{2,3} \simeq -0.425214 \pm 0.473898~i &~~~~ \rho_{2,3} \simeq
-1.41048 \pm 1.97212~i~,
\end{array}
\label{eqBRA}
\end{equation}
\renewcommand{\arraystretch}{1}
\noindent
cf.~refs.~\cite{JBKT,EHW}. The behavior of the real and imaginary part
of $\gamma_L(\rho)$, is illustrated in Fig.~1. For ${\rm Re}\, \rho
\rightarrow 4 \ln2$, the first branch point, ${\rm Re}~\gamma_L(\rho)$
forms a `roof' at $\gamma_L = 1/2$ for ${\rm Im}\, \rho = 0$, which
remains stable over some distance in ${\rm Re}~\rho$. The imaginary
part becomes discontinuous.
At smaller values of ${\rm Re}~\rho$, ${\rm Re}~\gamma_L(\rho)$
develops two symmetric minima, and for even smaller values two
additional maxima. Both extrema finally form the two other branch
points,~Eq.~(\ref{eqBRA}). In ${\rm Im}~\gamma_L$ these branch points
manifest as single extrema of the corresponding curve for ${\rm Re}
~\rho = {\sf const}\,$.

\subsection{Next-order corrections}
\label{sect32}

\noindent
The coefficients $\gamma_{qg,k}^{(1)}$ and $\gamma_{qq,k}^{(1)}$ of
the NL$x$ series in Eq.~(\ref{gres}) were calculated in ref.~\cite{CH}.
Recently also the first terms for $\gamma_{gg,k}^{(1)}$ have been
determined~\cite{CC1,CC2}. All these quantities are analytic functions
of $\gamma_L$, hence they do not introduce new singularities. The
energy-scale dependent contributions to $\gamma_{gg,k}^{(1)}$ have
still to be derived, and the terms $\gamma_{gq,k}^{(1)}$ are also
unknown so far in
\beq
\label{GAMNL}
 \GGV_{\rm NL}(N, \alpha_s) \equiv \ab \sum_{k=0}^{\infty} \left(
 \frac{\ab}{N} \right)^k \GGV_k^{(1)} = -2 \left( \begin{array}{cc}
 \!\! {\ds \frac{ C_F}{C_A} } [ \gamma_{qg, \rm NL} -
 \frac{\ds 2}{\ds 3 \pi} \alpha_s T_F ]  & \!\gamma_{qg, \rm NL} \! \\
 \!\gamma_{gq,{\rm NL}} & \!\gamma_{gg,{\rm NL}} \! \end{array}
 \right)_{\rm DIS} \: .
\eeq
In the DIS factorization scheme, the function $\gamma_{qg, \rm NL}
(N,\alpha_s)$ given by
\bea
\label{eqqg}
 \gamma _{qg, \rm NL}^{\rm DIS}(N, \alpha_s) &\! =\!&
 \gamma _{qg, \rm NL}^{Q_0}(N, \alpha_s) R(\gamma_L) \, =\,
 T_F \frac{\alpha_s}{6 \pi} \frac{2 + 3\gamma _L - 3\gamma _L^2}
 {3 - 2\gamma _L} \frac{[B(1- \gamma _L, 1+ \gamma _L)]^3}
 {B(2+ 2\gamma _L, 2- 2\gamma _L)} R( \gamma _L) \nonumber\\
 &\! \equiv \! &\frac{2 \alpha_s}{3 \pi} T_F \sum_{k=0}^{\infty}
 g_{k, qg}^{(1)} \left( \frac{\overline{\alpha}_s}{N} \right )^k
\label{eqS2}
\eea
with
\bea
 R(\gamma ) &\! =\! & \left[ \frac{\Gamma (1- \gamma ) \chi_0(\gamma )}
 {\Gamma (1+ \gamma ) \{ -\gamma \chi_0^{\prime}(\gamma )\}}\right]
 ^{1/2} \exp \left[ \gamma \psi (1) + \int_0^{\gamma }\! d\zeta \,
 \frac{\psi ^{\prime}(1) - \psi ^{\prime}(1 - \zeta )} {\chi_0 (\zeta )}
 \right] \equiv \sum_{k=0}^{\infty} r_k \left( \frac{\ab}{N} \right)^k~.
\nonumber\\
\label{rgam}
\eea
Here $B(x,y)$ denotes Euler's Beta-function, and $\gamma_{ij, \rm NL}
^{Q_0}$ represents the anomalous dimensions in the $Q_0$-scheme
\cite{CIA}, in which the factor $R(\gamma_L)$ does not appear.

The presently available contributions to the splitting function
$xP_{qg}(x)$ in the DIS scheme and their convolutions with a typical
gluon shape are shown in Fig.~2 for $\alpha_s = 0.2$, i.e., $Q^2
\simeq 20 \mbox{GeV}^2$. The LO splitting function vanishes like $x$
for $x \rightarrow 0$, their NLO counterpart is constant for $x
\rightarrow 0$. The strongly rising NL$x$ result \cite{CH} therefore
dominates below $x \sim 10^{-2}$. This dominance persists after the
convolution below $x \sim 10^{-3}$, although here the differences
are considerably smaller than for the splitting functions themselves.

The contributions
$\propto N_f$~[9,46--48]
and the
energy-scale independent terms $\propto C_A$~\cite{CC2,LIP2} of
$\gamma _{gg, \rm NL}(N, \alpha_s)$, cf. also~\cite{WHITE},
have been calculated recently.
As shown in ref.~\cite{CC1}, $\gamma_{gg, \rm NL}(N, \alpha_s)$ can be
obtained from the larger eigenvalue $\gamma_+$ of the resummed
anomalous dimension matrix
\beq
\gamma_{\pm} = \frac{1}{2} \left (\gamma_{qq} + \gamma_{gg} \right)
\pm \frac{1}{2} \sqrt{ \left (\gamma_{gg} - \gamma_{qq}\right)^2
+ 4 \gamma_{qg} \gamma_{gq}},
\eeq
where
\begin{eqnarray}
\gamma_{+} &\simeq& \gamma_{gg} + \frac{\gamma_{qg}\gamma_{gq}}
{\gamma_{gg} - \gamma_{qq}} = \gamma_{gg} + \frac{C_F}{C_A} \gamma_{qg}
+
O\left(\alpha_s^2
f(\ab /N)
\right) \: ,\nonumber\\
\gamma_{-} &\simeq& \gamma_{qq} - \frac{\gamma_{qg}\gamma_{gq}}
{\gamma_{gg} - \gamma_{qq}} = \gamma_{qq} - \frac{C_F}{C_A} \gamma_{qg}
+ O\left(\alpha_s^2 f(\ab /N)
\right) \: .
\label{eqapp}
\end{eqnarray}
These relations result from the fact that the quarkonic (upper-row)
entries in Eq.~(\ref{GAML}) vanish, unlike the gluonic ones.
Furthermore one has $\gamma_- \simeq - (C_F/C_A) \cdot (2 \alpha_s)/
(3 \pi) \cdot T_F$ due to Eq.~(\ref{GAMNL}). In the $Q_0$ scheme,
the present contributions to $\gamma_{+}$ are determined as the solution
of~\cite{CC1,CC2}
\begin{equation}
1 = \frac{\ab}{N} \left [ \chi_0(\gamma_+) + \alpha_s \chi_1(\gamma_+)
\right] \: ,
\end{equation}
which yields
\begin{equation}
\gamma_+^{(1)} = - \alpha_s \frac{\chi_1(\gamma_L)}
{\chi'_0(\gamma_L)}
\end{equation}
after a perturbative expansion. Finally $\chi_1(\gamma)$ reads
\begin{equation}
\chi_1(\gamma) = \chi_1^{q\overline{q},\rm a}(\gamma)
               + \chi_1^{q\overline{q},\rm na}(\gamma)
               + \chi_1^{gg}(\gamma),
\end{equation}
with
\begin{eqnarray}
\!\!\!
\alpha_s \chi_1^{q\overline{q}, \rm a}
&\!\! =\!\! & \frac{N_f \alpha_s}{\pi} \frac{C_F}{C_A}
\left( \frac{\pi}{\sin(\pi
 \gamma)} \right)^2 \frac{ \cos(\pi \gamma)}{3 - 2 \gamma} \frac{2 +
 3\gamma(1 - \gamma)}{(1 - 2\gamma)(1 + 2 \gamma)}
\nonumber
\\
\!\!\!
\alpha_s \chi_1^{q\overline{q}, \rm na}
&\!\! =\!\! & \frac{N_f \alpha_s}{6\pi}
 \left[ \frac{1}{2}
\left ( \chi_0^2(\gamma)  + \chi'_0(\gamma) \right)
 - \frac{5}{3} \chi_0(\gamma)
-  \left( \frac{\pi}{\sin(\pi
 \gamma)} \right)^2
\frac{ 3\cos(\pi \gamma)}{2(1 - 2 \gamma)} \frac{2 +
 3\gamma(1 - \gamma)}{(1 + 2\gamma)(3 - 2 \gamma)} \right ]
\nonumber\\
\!\!\!
\alpha_s \chi_1^{gg} &\!\! =\!\! & \frac{C_A \alpha_s}{4\pi}
 \left[ - \frac{11}{6} \left ( \chi_0^2(\gamma)  + \chi'_0(\gamma)
 \right) + \left (\frac{67}{9} - \frac{\pi^2}{3} \right) \chi_0(\gamma)
+ \left
 ( 6 \zeta(3) + \frac{\pi^2}{3\gamma(1 - \gamma)} + \tilde{h}(\gamma)
 \right) \right.  \nonumber\\ & &~~ \left.
 \quad\quad\mbox{}-\left( \frac{\pi}{\sin(\pi \gamma)} \right)^2
 \frac{\cos(\pi \gamma)}{3(1 - 2 \gamma)} \left ( 11 + \frac{\gamma(1
 - \gamma)} {(1 + 2\gamma)(3 - 2 \gamma)} \right) \right ] \: .
\label{eq5}
\end{eqnarray}
The function $\tilde{h}(\gamma)$ in eq.~(\ref{eq5}), which contributes
to $\gamma_{k,gg}^{(1)}$ only for $k \geq 2$, is given by
\beq
\label{eq6}
 \tilde{h}(\gamma) \simeq \sum_{l=1}^3 a_l \left( \frac{1}{l+\gamma} +
 \frac{1}{1+l + \gamma} \right)
\eeq
in approximate form, with $a_1 =0.72, a_2 = 0.28 $ and $a_3 = 0.16$
\cite{CC2}. From these results the $Q_0$-scheme anomalous dimension is
then inferred by \cite{CC1}
\begin{equation}
\tilde{\gamma}_{gg}^{(1),Q_0} =
\gamma_{gg}^{(1)} - \frac{\beta_0}{4 \pi} \alpha_s^2
\frac{d}{d \alpha_s}  \ln \left (
\gamma_{gg}^{(0)} \sqrt{-\chi_0'(\gamma_{gg}^{(0)})} \right)
= \gamma_+^{(1)} - \frac{C_F}{C_A} \gamma_{qg}^{(1)} \equiv
- \frac{\alpha_s\chi_1(\gamma_{gg}^{(0)})}{\chi_0'(\gamma_{gg}^{(0)})}~.
\end{equation}
Using the transformation of $\gamma_+$ to the DIS scheme~\cite{CC1},
and employing Eqs.~(\ref{eqqg}) and (\ref{eqapp}) as well, one finally
arrives at
\begin{eqnarray}
 \gamma_{gg,\rm NL}^{\rm DIS} &=& \gamma_{gg,\rm NL}^{Q_0}
 + \frac{\beta_0}{4\pi} \alpha_s^2 \frac{d \ln R(\gamma_L)}{d \alpha_s}
 + \frac{C_F}{C_A} \left[1-R(\gamma_L)\right]
\gamma_{qg,\rm NL}^{Q_0}
\nonumber\\
&\equiv& \frac{\alpha_s}{6 \pi} \sum_{k=0}^{\infty} \left[ N_f
 g_{k,gg}^{\, q\overline{q},(a)} + g_{k,gg}^{\, q\overline{q},(b)}
 + \Delta g_{k,gg}^{\, gg} \right ] \left(\frac{\ab}{N} \right)^{k}
\: .
\label{eq9}
\end{eqnarray}
The first terms for $\tilde{\gamma}_{gg,\rm NL}^{\, q\overline{q},
Q_0}$ read~:
\bea
\tilde{\gamma}_{gg,\rm NL}^{q\overline{q}, Q_0 }
&=&  - \frac{N_f \alpha_s}{6 \pi} \left \{
1 + \frac{23}{6}   \frac{\overline{\alpha}_s}{N}
+  \left [ \frac{71}{18} - \frac{\pi^2}{6} \right]
\left (\frac{\overline{\alpha}_s}{N} \right )^2
+ \left [ \frac{233}{27} - \frac{13}{36}\pi^2  - 8 \zeta(3)
\frac{C_F}{C_A} \right]
\left ( \frac{\overline{\alpha}_s}{N} \right )^3
\right.
\nonumber\\
& &~~~~~~~+ \left [ \frac{1276}{81}  - \frac{71}{108} \pi^2
+ \frac{79}{3} \zeta(3)
- \frac{7}{120} \pi^4
- \frac{52}{3} \zeta(3) \frac{C_F}{C_A} \right]
 \left ( \frac{\overline{\alpha}_s}{N} \right )^4  \nonumber\\
& &~~~~~~~+
\left [ \left(
\frac{8384}{243} - \frac{233}{162} \pi^2 + \frac{284}{9} \zeta(3)
- \frac{91}{720} \pi^4 + 2 \zeta(5) - \frac{4}{3} \zeta(3) \pi^2
\right ) \right.
\nonumber\\
& &~~~~~~~~~~+ \left. \left(\frac{4}{3} \zeta(3) \pi^2
- \frac{284}{9} \zeta3) - 16 \zeta(5)  \right )
\frac{C_F}{C_A} \right ]
\left ( \frac{\overline{\alpha}_s}{N} \right )^5
\nonumber\\
& &~~~~~~~+
\left[ \left( \left. \frac{45928}{729} - \frac{638}{243} \pi^2
- \frac{65}{18} \zeta(3) \pi^2 - 2 \zeta(3)^2 - \frac{497}{2160} \pi^4
+ \frac{125}{3} \zeta(5) + \frac{2330}{27} \zeta(3)
\right. \right. \right.
\nonumber\\
& &~~~~~~~~~~\left. \left. - \frac{31}{3024} \pi^6 \right ) + \left (
\frac{26}{9} \pi^2 \zeta(3) - \frac{104}{3} \zeta(5) - \frac{1864}{27}
\zeta(3) - 80 \zeta(3)^2 \right) \frac{C_F}{C_A} \right ]
 \left ( \frac{\overline{\alpha}_s}{N} \right )^6
\nonumber\\
& &~~~~~~~\left.
+ \, O \left ( \frac{\overline{\alpha}_s}{N} \right )^7
\right \}~.
\label{eqS3}
\eea
A similar expression can be derived for $\gamma_{gg,\rm  NL}^{\, gg,
Q_0}$. Because of the yet approximate expression for $\tilde{h}(\gamma)$
and the missing energy-scale dependent terms we will, however, only
list the numerical values of those still preliminary expansion
coefficients in Table~1.

In Fig.~3 the different approximations to the splitting function
$xP_{gg}(x)$ are displayed. Here both the LO and the NLO terms are
flat for $x \ra 0$, while the L$x$ contribution~\cite{LIPAT} causes a
strong rise as $x \ra 0$. Also shown are the presently known NL$x$
terms just discussed. The addition of the quarkonic (NL$x_
{q\overline{q}}$) contribution~\cite{CC1} reduces the resummation
effect almost down to the fixed-order results. The energy-scale
independent gluonic terms~\cite{CC2} have an even stronger impact,
in fact, they turn $xP_{gg}$ negative already at $x \sim 10^{-2}$ for
$\alpha_s \sim 0.2$. A similar, but milder pattern is observed for the
convolution $x(P_{gg} \otimes f)$ with a typical gluon shape which
illustrates the $Q^2$-slope of the gluon density induced by $P_{gg}$.
Note that the energy-scale dependent contributions to the NL$x$ terms
in $P_{gg}$ have still to be calculated. These terms or yet unknown
higher-order (NNL$x$) contributions may change the present behavior of
$P_{gg}(x)$.

\begin{table}\centering
\begin{tabular}{||r||r|r|r|r|r||}
\hline\hline
& & & & & \\[-4mm]
\multicolumn{1}{||c||}{$k$} &
\multicolumn{1}{c|}{$g_{k, gg}^{(0)}$} &
\multicolumn{1}{c|}{$g_{k, qg}^{(1)}$ ($Q_0$)} &
\multicolumn{1}{c|}{$g_{k, qg}^{(1)}$ (DIS)}   &
\multicolumn{1}{c|}{$r_k$} &
\multicolumn{1}{c||}{$c^{\, L}_k$} \\
& & & & & \\[-4mm] \hline\hline & & & & & \\[-4mm]
  0 &  1.00000$\,$E+0   &  1.00000$\,$E+0   &  1.00000$\,$E+0  &
       1.00000$\,$E+0   &  1.00000$\,$E+0   \\
  1 &  0.00000$\,$E+0   &  2.16667$\,$E+0   &  2.16667$\,$E+0  &
       0.00000$\,$E+0   &--3.33333$\,$E$-$1 \\
  2 &  0.00000$\,$E+0   &  2.29951$\,$E+0   &  2.29951$\,$E+0  &
       0.00000$\,$E+0   &  2.13284$\,$E+0   \\
  3 &  2.40411$\,$E+1   &  5.06561$\,$E+0   &  8.27109$\,$E+0  &
       3.20549$\,$E+0   &  2.27231$\,$E+0   \\
  4 &  0.00000$\,$E+0   &  8.79145$\,$E+0   &  1.49249$\,$E+1  &
     --8.11742$\,$E$-$1 &  4.34344$\,$E$-$1 \\
  5 &  2.07386$\,$E+1   &  1.90521$\,$E+1   &  2.92268$\,$E+1  &
       4.56248$\,$E+1   &  2.02643$\,$E+1   \\
  6 &  1.73393$\,$E+1   &  4.58482$\,$E+1   &  1.02812$\,$E+2  &
       3.27070$\,$E+1   &  2.30315$\,$E+1   \\
  7 &  2.01670$\,$E+0   &  9.24159$\,$E+1   &  1.94887$\,$E+2  &
     --2.95476$\,$E+1   &  3.46449$\,$E+1   \\
  8 &  3.98863$\,$E+1   &  2.31063$\,$E+2   &  4.85100$\,$E+2  &
       1.08183$\,$E+2   &  2.65004$\,$E+2   \\
  9 &  1.68747$\,$E+2   &  5.59958$\,$E+2   &  1.52444$\,$E+3  &
       3.99588$\,$E+2   &  3.30038$\,$E+2   \\
 10 &  6.99881$\,$E+1   &  1.24822$\,$E+3   &  3.11451$\,$E+3  &
       1.33228$\,$E+2   &  8.50371$\,$E+2   \\
 11 &  6.61253$\,$E+2   &  3.25381$\,$E+3   &  8.58375$\,$E+3  &
       2.10243$\,$E+3   &  3.90849$\,$E+3   \\
 12 &  1.94531$\,$E+3   &  7.93653$\,$E+3   &  2.47571$\,$E+4  &
       5.51142$\,$E+3   &  5.67433$\,$E+3   \\
 13 &  1.71768$\,$E+3   &  1.89275$\,$E+4   &  5.47435$\,$E+4  &
       5.30316$\,$E+3   &  1.77680$\,$E+4   \\
 14 &  1.06433$\,$E+4   &  4.98520$\,$E+4   &  1.56195$\,$E+5  &
       3.85296$\,$E+4   &  6.21982$\,$E+4   \\
 15 &  2.55668$\,$E+4   &  1.23011$\,$E+5   &  4.26980$\,$E+5  &
       8.49086$\,$E+4   &  1.07028$\,$E+5   \\
 16 &  3.67813$\,$E+4   &  3.06504$\,$E+5   &  1.01111$\,$E+6  &
       1.40384$\,$E+5   &  3.51475$\,$E+5   \\
 17 &  1.71685$\,$E+5   &  8.07771$\,$E+5   &  2.89398$\,$E+6  &
       6.94998$\,$E+5   &  1.05058$\,$E+6   \\
 18 &  3.75379$\,$E+5   &  2.02210$\,$E+6   &  7.69042$\,$E+6  &
       1.44307$\,$E+6   &  2.10341$\,$E+6   \\
 19 &  7.36025$\,$E+5   &  5.17873$\,$E+6   &  1.91919$\,$E+7  &
       3.22738$\,$E+6   &  6.80747$\,$E+6   \\[1mm]
\hline\hline
& & & & & \\[-4mm]
\multicolumn{1}{||c||}{$k$} &
\multicolumn{1}{c|}{$g_{k,gg}^{\, q\overline{q}\, (a)}$ (Q$_0)$} &
\multicolumn{1}{c|}{$g_{k,gg}^{\, q\overline{q}\, (b)}$ (Q$_0)$} &
\multicolumn{1}{c|}{$g_{k,gg}^{\, q\overline{q}\, (a)}$ (DIS)} &
\multicolumn{1}{c|}{$g_{k,gg}^{\, q\overline{q}\, (b)}$ (DIS)} &
\multicolumn{1}{c||}{$\Delta g_{k,gg}^{\, gg}$} \\
& & & & & \\[-4mm] \hline\hline & & & & & \\[-4mm]
  0 &--1.00000$\,$E+0   &  0.00000$\,$E+0   &--1.00000$\,$E+0   &
       0.00000$\,$E+0   &--1.65000$\,$E+1   \\
  1 &--3.83333$\,$E+0   &  0.00000$\,$E+0   &--3.83333$\,$E+0   &
       0.00000$\,$E+0   &  0.00000$\,$E+0   \\
  2 &--2.29951$\,$E+0   &  0.00000$\,$E+0   &--2.29951$\,$E+0   &
       0.00000$\,$E+0   &  1.48980$\,$E+1   \\
  3 &  6.42072$\,$E+0   &--1.19004$\,$E+2   &--6.04506$\,$E+0   &
       3.96679$\,$E+1   &--2.25291$\,$E+2   \\
  4 &--2.59764$\,$E+1   &  0.00000$\,$E+0   &--2.81814$\,$E+1   &
     --5.35750$\,$E+1   &  2.42631$\,$E+0   \\
  5 &  5.75787$\,$E+0   &--3.42186$\,$E+2   &--2.60988$\,$E+1   &
       3.42186$\,$E+1   &--2.09481$\,$E+2   \\
  6 &  1.21690$\,$E+2   &--2.28879$\,$E+3   &--9.43607$\,$E+1   &
       4.40583$\,$E+2   &--2.78211$\,$E+3   \\
  7 &--2.66365$\,$E+2   &--6.98786$\,$E+2   &--3.54981$\,$E+2   &
     --7.39527$\,$E+2   &--2.70970$\,$E+2   \\
  8 &  5.43807$\,$E+2   &--1.11881$\,$E+4   &--4.27828$\,$E+2   &
       1.11801$\,$E+3   &--7.53012$\,$E+3   \\
  9 &  1.96852$\,$E+3   &--4.10835$\,$E+4   &--1.67366$\,$E+3   &
       4.86665$\,$E+3   &--3.82351$\,$E+4   \\
 10 &--2.04998$\,$E+3   &--3.39345$\,$E+4   &--5.21390$\,$E+3   &
     --9.10195$\,$E+3   &--1.56381$\,$E+4   \\
 11 &  1.49302$\,$E+4   &--2.75933$\,$E+5   &--7.99079$\,$E+3   &
       2.40902$\,$E+4   &--1.73122$\,$E+5   \\
 12 &  3.33837$\,$E+4   &--7.55104$\,$E+5   &--3.05607$\,$E+4   &
       5.32758$\,$E+4   &--5.68231$\,$E+5   \\
 13 &  9.19579$\,$E+3   &--1.10387$\,$E+6   &--8.37332$\,$E+4   &
     --9.58437$\,$E+4   &--5.13392$\,$E+5   \\
 14 &  3.35804$\,$E+5   &--6.12763$\,$E+6   &--1.57171$\,$E+5   &
       4.46747$\,$E+5   &--3.52577$\,$E+6   \\
 15 &  6.26484$\,$E+5   &--1.45966$\,$E+7   &--5.64262$\,$E+5   &
       5.92510$\,$E+5   &--9.13527$\,$E+6   \\
 16 &  9.72892$\,$E+5   &--3.01102$\,$E+7   &--1.43675$\,$E+6   &
     --6.85258$\,$E+5   &--1.36171$\,$E+7   \\
 17 &  7.05626$\,$E+6   &--1.30018$\,$E+8   &--3.14592$\,$E+6   &
       7.71985$\,$E+6   &--6.85495$\,$E+7   \\
 18 &  1.29507$\,$E+7   &--2.96814$\,$E+8   &--1.05144$\,$E+7   &
       7.22515$\,$E+6   &--1.58648$\,$E+8   \\
 19 &  3.18568$\,$E+7   &--7.45406$\,$E+8   &--2.59548$\,$E+7   &
       2.95797$\,$E+6   &--3.23625$\,$E+8   \\[1mm]
\hline\hline
\end{tabular}
 \caption[XX]{\sf The numerical expansion coefficients in Eqs.~(\ref
 {GAML}), (\ref{eqS2}), (\ref{rgam}), (\ref{eq9}) and (\ref{eqS4}).}
\end{table}

The leading singular contributions to the gluonic and pure-singlet
quarkonic coefficient functions for the longitudinal structure
function were also determined in ref.~\cite{CH},
\bea
\label{eqS4}
C_L^g &\! =\! & \frac{ \alpha_s}{3 \pi}  T_F  \left (
\frac{1 - \gamma_L} {3 - 2 \gamma_L} \right)
\frac{\left [ B(1 - \gamma_L, 1 + \gamma_L) \right ]^3 }
{B(2 - 2 \gamma_L, 2 + 2 \gamma_L)} R(\gamma_L)
\, \equiv \,  \frac{2 \alpha_s}{3 \pi} T_F
\sum_{k=1}^{\infty} c_k^{\, L} \left (
\frac{\overline{\alpha}_s}{N} \right )^k, \\
\label{eqS5}
C_L^S &\! = \! & \frac{C_F}{C_A} \left [ C_L^g -
\frac{2 \alpha_s}{3 \pi} T_F \right ]~.
\eea
The numerical values of the first 20 expansion coefficients
$g_{k,gg}^{(0)}, g_{k,qg}^{(1)}$ in the DIS and $Q_0$ schemes,
$r_k, c_k^{\, L}$, and coefficients contributing to
$g_{k,gg}^{\,q\overline{q}}$ and $\Delta g_{k,gg}^{\, gg}$ are listed
in Table~1 for completeness. These coefficients were calculated
using the {\tt MAPLE}-package~\cite{MAPLE}. The numerical values of
$g_{k,qg}^{(1)}/(4 \ln 2)^k$ were tabulated before for the DIS
\cite{BF1} and $\,\overline{\rm MS}$ schemes~\cite{BF2}. With a low
number of digits the values of $g_{k,qg}^{\overline{\rm MS},\, (1)}$,
$r_k$ and $c_k^L$ were given in ref.~\cite{FRT} as well. Either the 
direct expressions~(\ref{GAML}), (\ref{eqqg}) or relations based on 
the corresponding expansion coefficients have been previously used in
numerical studies~\cite{EHW,BRV,STU3}.

\subsection{Photon-parton splitting functions}
\label{sect33}

\noindent
Finally we have to consider the small-$x$ higher-order corrections
to the inhomogeneous terms $k_q \equiv P_{q\gamma}$ and $k_g \equiv 
P_{g\gamma}$ in the photonic evolution equations (\ref{evolx}). These
quantities arise from a subset of the diagrams leading to the 
gluon-parton splitting functions, with the incoming gluon replaced by 
a photon. Purely gluonic graphs do obviously not belong to that 
subset, as the photon can couple to the hadronic system only through 
$q\bar{q}$ emission. Hence, by comparing to the hadronic results
discussed above, one obtains the most singular $(S)$ small-$x$ terms as
\beq
  k_q^{\, S} = a_{\rm em}\, k_{q,0} (N\! =\! 0) + \sum_{l=1}^{\infty} 
  a_{\rm em} a_s^l \, \frac{k_{q,0}^{\, S}}{N^{l-1}} \:\: , \:\:\:
  k_g^{\, S} = \sum_{l=1} ^{\infty} a_{\rm em} a_s^l \,\frac{k_{g,0}
 ^{\, S}}{N^l} \: . 
\eeq
I.e., $k_q$ (up to its scheme-independent constant LO term) is 
definitely beyond the current NL$x$ approximation, whereas $k_g$ can 
receive contributions at this order. Likewise, the constant but 
scheme-dependent term $C_{\gamma ,1}$ is NL$x$, and all higher-order 
photonic coefficient functions are beyond that approximation. A 
$1/N$-term is indeed present in the known NLO result $k_{g,1}$. This 
term, however, vanishes after transformation to the DIS or 
DIS$_{\gamma}$ schemes. In fact, this cancellation can always be 
achieved at all orders, as we will show now.

We proceed in two steps. We start by a purely hadronic transformation
(like that one from \MSbar to DIS), where $\CV \rightarrow 
\widetilde{\CV} = \CV + \Delta \CV$, and $\CV_{\gamma}$ remains 
unchanged. Then Eq.~(\ref{sinv}) implies
\beq
  0 = \Delta \IV_{\rm inh.} = \widetilde{\CV}\cdot \Delta \kV + 
  \Delta \CV \cdot \kV \: .
\eeq
We consider only such transformation terms, which are motivated in
contributions to the actual electromagnetic coefficient functions.
These are, however, of NL$x$ order, as the $\kV$ themselves, and hence
one has $\Delta \kV = 0$ on the NL$x$ level: purely hadronic scheme
changes modify the photon-parton splitting functions only beyond
the NL$x$ approximation.

The second step is a purely photonic transformation, $\CV_{\gamma}
\rightarrow \widetilde{\CV}_{\gamma} = \CV_{\gamma} + \delta 
\CV_{\gamma}$, with $\CV $ untouched. Here Eq.~(\ref{sinv}) yields
\beq
  0 = \Delta \IV_{\rm inh.} = \CV \cdot \Delta \kV + \beta \cdot \Delta
  \CV_{\gamma}^{\prime} - \IV_{\rm hom.} \cdot \Delta \CV_{\gamma} \: .
\eeq
Solving for the changes $\Delta \kV_m$ up to the order $a_{\rm em} 
a_s^m$ therefore leads to
\bea
 \Delta \kV_1 &=& \: \IV_{\rm hom., 0} \cdot \Delta \CV_{\gamma ,1}
  \, \nonumber \\
 \Delta \kV_2 &=& - \CV_1 \cdot \Delta \kV_1 + \IV_{\rm hom., 0} 
  \cdot \Delta \CV_{\gamma ,2} + \IV_{\rm hom., 1} \cdot \Delta 
  \CV_{\gamma ,1} \, \\
  &\vdots & \nonumber\\
 \Delta \kV_m &=& - \sum_{l=1}^{m-1} \CV_l \cdot \Delta \kV_{m-l}
  + \sum_{l=0}^{m-1} \IV_{\rm hom., l} \cdot\Delta \CV_{\gamma ,m-l+1}
  \: \nonumber ,
\eea
where only terms have been retained which can potentially contribute to
$\Delta \kV_{{\rm NL}x}$, if an NL$x$ contribution to $\Delta 
\CV_{\gamma}$ occurs in the transformation. By choice of the 
(unphysical) lower component of $\Delta \CV_{\gamma ,m}$ the NL$x$
pieces of $\kV_m$, arising, for example, in an \MSbar calculation, can 
hence successively be eliminated, without disturbing the vanishing of 
the upper term.  At NLO, e.g., the lower component of zero in Eq.\ 
(\ref{Ctrf}), as chosen in the DIS$_{\gamma}$ scheme \cite{GRVph1}
achieves this cancellation. Therefore, without any loss of generality, 
vanishing resummed photon-parton splitting functions can be assumed at 
the NL$x$ level.

\section{Less singular contributions}
\label{sect4}
\setcounter{equation}{0}

\noindent
The terms in the splitting functions $P_{ij}$, which are less singular
by one (or more) powers of $\ln (1/x)$ as $x \ra 0 $ than the leading
contributions discussed in the previous section, are presently unknown
in almost all cases. Such subleading contributions, however, can
potentially prove to be as important as the leading terms, as also
noted in a similar context in ref.~\cite{SUBL}. The splitting functions
and coefficient functions enter observable quantities always via Mellin
convolutions with the parton distributions. Since the parton densities
are steeply rising towards small-$x$, but (at least in the hadronic
case) small at large $x$, the structure functions probe the behavior of
splitting functions and coefficient functions at medium and large values
of $x$ as well.

The unpolarized singlet splitting functions are constrained by
energy-momentum conservation, see Eq.~(\ref{conserv}).
Also in other cases, however, as for the polarized
singlet and the non-singlet-$+$ evolutions, where no conservation laws
apply, less singular terms with sizeable coefficients exist, for
example in NLO, see, e.g., ref.~\cite{BRV}. In order to evaluate the
possible impact of such terms in higher-order splitting functions,
their numerical coefficients need to be estimated. At present the
almost only source of information are the fully known LO and NLO
splitting functions.

The dominant and subdominant terms in the small-$x$ $1/N$-expansion of
the LO and NLO singlet anomalous dimensions are recalled in Eqs.\
(\ref{eqSU1})--(\ref{eqSU3}). In accordance with the main part of our
numerical studies in Sect.~6, the results are listed for four light
quark flavors.  The first terms in the LO case are given by
\begin{eqnarray}
\label{eqSU1}
\gamma_{qq, \rm LO} &=&  + 10.8793\, N - 6.82222\, N^2 + O(N^3)
                         \: ,  \nonumber\\
\gamma_{qg, \rm LO} &=&  - 10.6667   + 11.5556\, N - 13.1852\, N^2
                         + O(N^3) \: , \nonumber\\
\gamma_{gq, \rm LO} &=&  - \frac{10.6667}{N}  + 8.00000  - 9.3333\, N
                        + 10.0000\, N^2 + O(N^3) \: , \nonumber\\
\gamma_{gg, \rm LO} &=&  - \frac{24.0000}{N}  + 27.3333 - 5.1883\, N
                        + 17.0395\, N^2 + O(N^3) \: .
\end{eqnarray}
The corresponding expansions of the NLO anomalous dimensions read, in
the DIS scheme,
\begin{eqnarray}
\gamma_{qq, \rm NLO}^{\rm DIS} &=&  - \frac{123.259}{N} + 405.863
          - 684.836\, N + 1197.52 N^2 + O(N^3) \: , \nonumber\\
\gamma_{qg, \rm NLO}^{\rm DIS} &=&   - \frac{277.333}{N} + 846.222
          - 1706.18\, N + 2622.76\, N^2 + O(N^3) \: , \nonumber\\
\gamma_{gq, \rm NLO}^{\rm DIS} &=&  + \frac{91.2593}{N} - 453.512
          + 809.030\, N - 1344.89\, N^2 + O(N^3) \: , \nonumber\\
\gamma_{gg, \rm NLO}^{\rm DIS} &=&  + \frac{245.333}{N} - 988.210
          + 2093.25\, N  - 3109.08\, N^2 + O(N^3) \: .
\label{eqSU2}
\end{eqnarray}

One notices that the first subleading terms occur  with a sign opposite
to that of the dominant one. Their prefactors are of the same order,
but in most cases the subleading coefficients are by a factor
of about 2 to
4 larger. At leading order the quarkonic terms are not singular as
$N \ra 0$. The $qq-$term even starts proportional to $N$, as a
consequence of fermion-number conservation, Eq.~(\ref{fcons}). The
alternating structure continues towards higher powers in $N$ with a
similar pattern for the coefficients as observed for the first and
second term. Note that this behavior is not a special feature of the
DIS scheme, but is observed to a similar extent also in other schemes.
As an example we give the corresponding coefficients also for the
$\overline{\rm MS}$-scheme~:
\begin{eqnarray}
\gamma_{qq, \rm NLO}^{\overline{\rm MS}}
&=& - \frac{94.8148}{N}  + 253.026 - 337.185\, N + 623.259\, N^2
                                    + O(N^3) \: , \nonumber\\
\gamma_{qg, \rm NLO}^{\overline{\rm MS}}
&=& - \frac{213.333}{N}  + 461.449 - 889.687\, N + 1501.16\, N^2
                                    + O(N^3) \: , \nonumber\\
\gamma_{gq, \rm NLO}^{\overline{\rm MS}}
&=& + \frac{62.8148}{N}  - 361.805 + 658.108\, N - 1048.43\, N^2
                                    + O(N^3) \: , \nonumber\\
\gamma_{gq, \rm NLO}^{\overline{\rm MS}}
&=& + \frac{216.889}{N}  - 790.928 + 1616.55\, N - 2423.77\, N^2
                                    + O(N^3) \: .
\label{eqSU3}
\end{eqnarray}
In the structure function evolution, the difference between Eqs.\
(\ref{eqSU2}) and (\ref{eqSU3}) is compensated by the corresponding
small-$x$ terms of the coefficient functions.

In the small-$x$ resummation case, even partial results for subdominant
contributions are only available for the gluon-gluon splitting
function so far. The irreducible NL$x$ contributions to $\gamma_{gg}$
\cite{CC2} exhibit very large coefficients if compared to the L$x$
series \cite{LIPAT}, see Table~1 and the comparison in ref.~\cite
{BV97}. The  introduction of terms with prefactors up to two times
larger than those of the leading contributions, therefore, should
provide conservative, non-exaggerating estimates for the possible
impact of subdominant corrections.
The following modifications of the resummed anomalous dimensions
beyond two-loop order, $\Gamma(N,\alpha_s)$, have accordingly been
studied within refs.~\cite{BVplb1,BRV,EHW,BVplb2}:
\begin{equation}
\label{xxx}
\begin{array}{cl}
{\rm (A):} & \Gamma(N, \alpha_s)
\rightarrow  \Gamma(N, \alpha_s) - \Gamma(1, \alpha_s)
\\[1mm]
{\rm (B):} & \Gamma(N, \alpha_s)
\rightarrow  \Gamma(N, \alpha_s)(1 - N)
\\[1mm]
{\rm (C):} & \Gamma(N, \alpha_s)
\rightarrow  \Gamma(N, \alpha_s)
(1 - N)^2
\\[1mm]
{\rm (D):} & \Gamma(N, \alpha_s)
\rightarrow  \Gamma(N, \alpha_s)(1 - 2 N + N^3) .
\end{array}
\end{equation}

The impact of the prescriptions (C) and (D) on the resummed NL$x$
contribution to the splitting function $xP_{qg}(x,\alpha_s)$ is
illustrated in Fig.~2. At $x = 10^{-4}$ those terms reduce $xP_{qg}$
by factors larger than three, indicating the potential importance of
less singular contributions. Also displayed in Fig.~2 is the
convolution of $P_{qg}$ with a typical hadronic gluon shape. The
enhancement of the importance of non-leading terms by the Mellin
convolution discussed above is obvious from the comparison of the
two plots.
\section{Solution of the evolution equations}
\label{sect5}
\setcounter{equation}{0}

\noindent
In this section we derive the solution of the singlet evolution
equations presented above. For technical convenience the analysis is
performed in Mellin-$N$ space where the convolutions turn to simple
products. Recall that a unique analytic continuation of the anomalous
dimensions from the integer moments to complex $N$ exists~\cite{UNIQ}.
Thus a coupled system of two ordinary differential equations has to be
solved at fixed $N$. The $x$-space results are then obtained by a
contour integral around the singularities of the final moment
solutions $f(N)$ in the complex $N$-plane, e.g., that shown in Fig.~4.
Due to $f^*(N) = f(N^*)$ it yields~\cite{GRV90}
\beq
\label{minv}
 xf(x) = \frac{1}{\pi} \int^{\infty}_0 \!\! dz \, {\rm Im}
 [ e^{i \phi} x^{-C} f(N\! =\! C) ] \: ,
\eeq
where $C = c + z e^{i \phi} $.  For all cases considered here,
$ c \simeq 1 $ and $\phi \simeq 3\pi /4 $ provide an efficient and
numerically stable inversion. The latter choice of $\phi > \pi /2$
leads to a faster convergence of the integral (\ref{minv}) as
$z \rightarrow \infty$, see also ref.~\cite{BGHV}. At small-$x$, for
example, a numerical accuracy better than $10^{-5}$ is easily achieved
for upper limits as low as $z_{\rm max} \simeq 5$.

\subsection{The general hadronic solution}
\label{sect51}

\noindent
It is convenient to recast the evolution equations in terms of the
running coupling $ a_s = \alpha_s(Q^2)/4\pi $ as independent variable,
by combining the $Q^2$ evolution (\ref{evol1}) of the hadronic parton
densities $\qV$ with Eq.~(\ref{eval}) for $a_s$. Sorting the resulting
r.h.s.\ in powers of $a_s$, one obtains
\bea
\label{evol2a}
 \frac{\partial \qV (a_s,N)}{\partial a_s}
 &=& \frac{a_s \PV_0(N) + a_s^2 \PV_1(N) + a_s^3 \PV_2(N)+ \ldots}
  {-a_s^2 \,\beta_0 - a_s^3 \,\beta_1 - a_s^4 \,\beta_2 - \ldots}
  \: \qV (a_s,N) \nonumber\\
 &=& -\frac{1}{\beta_0 a_s}
  \bigg[ \PV_0 (N) + a_s \bigg( \PV_1 (N)- \frac{\beta_1}{\beta_0}
  \PV_0 (N)\bigg) \nonumber \\ & & \mbox{} \hspace*{10mm}
  + a^2_s \bigg( \PV_2 (N) - \frac{\beta_1}{\beta_0} \PV_1 (N)
  + \bigg\{ \bigg( \frac{\beta_1}{\beta_0} \bigg)^2 - \frac{\beta_2}
  {\beta_0} \bigg\} \PV_0 (N) \bigg) + \ldots \bigg]
  \, \qV (a_s,N) \nonumber \\
 & = & -\frac{1}{a_s}
  \bigg[ \RV_0 (N) + \sum_{k=1}^{\infty} a_s^k \RV_k (N) \bigg]
  \, \qV (a_s,N) \: .
\eea
Here we have simplified the notation by introducing the recursive
abbreviations
\bea
\label{r0}
  \RV_0 &\equiv & \frac{1}{\beta_0} \PV_0 \: , \\
  \RV_k &\equiv & \frac{1}{\beta_0} \PV_k - \sum_{i=1}^{k}
                  \frac{\beta_{i}} {\beta_0} \RV_{k-i}
\label{rmat}
\eea
for the splitting function combinations entering this expansion. As in
Eqs.~(\ref{r0}) and (\ref{rmat}), we will often suppress the explicit
reference to the Mellin variable $N$ below.

The splitting function matrices $\PV_k $ of different orders $k$ do
generally not commute, especially one has
\beq
  \big[ \RV_{k \geq 1}(N), \RV_0(N) \big] \neq 0 \: .
\eeq
This prevents, already at NLO, writing the solution of Eq.\
(\ref{evol2a}) in a closed exponential form. Instead we proceed by
generalizing the NLO method of ref.~\cite{FP} to all orders$\,
$\footnote {The first three orders were treated in a very similar
manner in ref.~\cite{EKL}.} in $a_s$. The corresponding ansatz of a
series expansion around the lowest order solution,
\beq
\label{sol0}
 \qV^{\,\rm LO} (a_s,N) =
 \left( \frac{a_s}{a_0} \right)^{{\footnotesize -\RV}_0 (N)}
 \qV (a_0,N) \equiv \LV (a_s,a_0,N) \, \qV (a_0,N) \: ,
\eeq
reads
\bea
\label{sol1}
 \qV (a_s,N) &=& \UV (a_s,N) \LV (a_s,a_0,N) \UV^{-1}(a_0, N) \:
 \qV (a_0,N) \\
 &=& \Big[ 1 + \sum_{k=1}^{\infty} a^k_s\UV_k (N) \Big] \LV (a_s,a_0,N)
 \Big[ 1 + \sum_{k=1}^{\infty} a_0^k \UV_k (N) \Big]^{-1} \qV (a_0,N)
 \: . \nonumber
\eea
The third, $a_s$-independent factor in Eq.~(\ref{sol1}) has been
introduced to normalize the evolution operator to the unit matrix at
$Q_0^2$, instead of to the LO result (\ref{sol0}) at infinitely high
$Q^2 $.
Inserting this ansatz into the evolution equations (\ref{evol2a}) and
sorting in powers of $a_s$ anew, one arrives at a chain of commutation
relations for the expansion coefficients $\UV_k(N) $:
\bea
\label{Ueq}
  \big[ \UV_1, \RV_0 \big] & = & \RV_1 + \UV_1 \: , \nonumber \\
  \big[ \UV_2, \RV_0 \big] & = & \RV_2 + \RV_1 \UV_1 +
            2 \UV_2 \: , \\
  & \vdots & \nonumber \\
  \big[ \UV_k, \RV_0 \big] & = & \RV_k + \sum_{i=1}^{k-1}
            \RV_{k-i} \UV_i + k \UV_k
  \: \equiv \: \widetilde{\RV}_k + k \UV_k \: . \nonumber
\eea

These equations can be solved recursively by applying the eigenvalue
decomposition of the LO splitting function matrix, completely analogous
to the truncated NLO solution with only $\UV_1 $ \cite{FP}, see below.
One writes
\beq
  \RV_0 = r_{-} \eV_{-} + r_{+} \eV_{+} \: ,
\eeq
where $r_{-}$ ($r_{+}$) stands for the smaller (larger) eigenvalue of
$ \RV_0 $,
\beq
  r_{\pm } = \frac{1}{2 \beta_0} \bigg[ P_{qq}^{(0)} + P_{gg}^{(0)}
  \pm \sqrt{ \Big( P_{qq}^{(0)} - P_{gg}^{(0)}\Big) ^2 +
  4 P_{qg}^{(0)} P_{gq}^{(0)} } \, \bigg] \: .
\eeq
The matrices $\eV_{\pm }$ denote the corresponding projectors,
\beq
 \eV_{\pm} = \frac{1}{r_{\pm}-r_{\mp}} \Big[ \RV_0-r_{\mp}\IV \Big]\: ,
\eeq
with $\IV $ being the $2\!\times\! 2 $ unit matrix. Hence the LO
evolution operator (\ref{sol0}) can be represented as
\beq
 \LV (a_s,a_0,N) = \eV_{-}(N) \bigg(\frac{a_s}{a_0}\bigg)^{-r_{-}(N)}
   + \eV_{+}(N) \bigg(\frac{a_s}{a_0}\bigg)^{-r_{+}(N)} \: .
\eeq
Inserting the identity
\beq
  \UV_k = \eV_{-} \UV_k \eV_{-} + \eV_{-} \UV_k \eV_{+} +
          \eV_{+} \UV_k \eV_{-} + \eV_{+} \UV_k \eV_{+}
\eeq
into the commutation relations (\ref{Ueq}), one finally obtains the
expansion coefficients in Eq.\ (\ref{sol1}):
\beq
\label{Usol}
  \UV_k =
  - \frac{1}{k} \Big[ \eV_{-} \widetilde{\RV}_k \eV_{-} +
    \eV_{+} \widetilde{\RV}_k \eV_{+} \Big]
  + \frac{\eV_{+} \widetilde{\RV}_k \eV_{-}}{r_{-} - r_{+} - k}
  + \frac{\eV_{-} \widetilde{\RV}_k \eV_{+}}{r_{+} - r_{-} - k} \: .
\eeq
This relation completes the general structure of the hadronic singlet
evolution. Note that the poles in $\UV_k (N)$ at $N$-values where
$r_{-}(N) - r_{+}(N) \pm k $ vanishes are canceled by the $\UV^{-1} $
term in the solution (\ref{sol1}). We are now ready to consider the
presently available fixed-order and small-$x$ resummation
approximations.

\subsection{Fixed-order evolution}
\label{sect52}

In fixed-order perturbative QCD the expansion (\ref{Ps}) of the
splitting functions in powers of the strong coupling $a_s$ is truncated
at some low order $k$. Practical small-$x$ calculations are presently
restricted to NLO ($k=1$), as the NNLO splitting functions $\PV_2(N)$
are not yet known for arbitrary values of $N$, unlike the 2-loop
coefficient functions \cite{CO2,CO21} and the $\beta $-function
coefficient $\beta_2$ \cite{beta2}. Hence we confine ourselves to the
NLO evolution here, the generalization to higher fixed orders being
obvious. I.e., we keep the full results up to $k=1$ and put in
Eq.~(\ref{evol2a})
\bea
\label{PNLO}
 \PV_{k \geq 2}(N) &=& 0 \: , \\
 \beta_{k \geq 2}  &=& 0 \: .
\label{bNLO}
\eea
The coefficients $\beta_{k \geq 2}$ are also removed, for only all
three quantities $\PV_k$, $C_k$ and $\beta_k$ together form a scheme
independent set for the evolution of physical quantities like the
structure functions $F_2(x,Q^2)$ or, in the case of polarized
scattering, $g_1(x,Q^2)$.

Two natural approaches have been widely adopted for the solution of the
resulting NLO evolution equations. First one can solve
Eq.~(\ref{evol2a}) as it stands after inserting Eqs.\ (\ref{PNLO}) and
(\ref{bNLO}). Then still all orders in $a_s$ contribute there and in
the solution (\ref{sol1}), with the only simplification that the
splitting function combinations (\ref{rmat}) are now given by
\beq
\label{rnlo1}
 \RV_k^{\,\rm NLO} = \frac{(-1)^{k-1}}{\beta_0} \bigg( \frac{\beta_{1}}
 {\beta_0} \bigg)^{k-1} \Big( \PV_1 - \frac{\beta_{1}}{\beta_0} \PV_{0}
 \Big) \: .
\eeq
This procedure is equivalent to a simple iterative solution of the
system (\ref{evol1}) and (\ref{eval}), truncated at $k=1$. That
technique is widely used in parton density analyses, e.g., in
refs.~[60--62].

The second approach uses power counting in $a_s$ at the level
of the evolution equation~(\ref{evol2a}). There the $a_s^2$ term in the
square bracket involves $\PV_2 $ and $\beta_2$ and can thus by
considered as beyond the present approximation. Consequently only the
constant and the linear terms in $a_s$ are kept, instead of Eq.\
(\ref{rnlo1}) leading to
\beq
\label{rnlo2}
 \RV_1^{\,\rm NLO'} = \RV_1^{\,\rm NLO} = \frac{1}{\beta_0}
 \Big( \PV_1 - \frac{\beta_{1}} {\beta_0} \PV_{0} \Big) \: , \:\:\:\:\:
 \RV_{k \geq 2}^{\,\rm NLO'} = 0 \: .
\eeq
In this approach it is furthermore natural to truncate also the
evolution matrix $\UV (a_s)$ after the linear term, since $\PV_2 $
would enter the determination of $\UV_2 $ in Eq.~(\ref{Ueq}) as well.
Recall also that the final multiplication with the NLO Wilson
coefficients does only cancel the scheme dependence of the linear $a_s$
term in the evolution of the structure functions (\ref{strf}). Finally
one can expand also $\UV^{-1}(a_0)$ to first order in Eq.~(\ref{sol1}),
although this is not necessary, resulting in
\beq
\label{sol2}
 \qV^{\,\rm NLO}_{\,\rm tr.} (a_s,N) = \Big[ \LV (a_s,a_0,N) + a_s
 \UV_1(N) \LV (a_s,a_0,N) - a_0 \LV (a_s,a_0,N) \UV_1(N) \Big] \:
\qV (a_0,N) \: ,
\eeq
where $\UV_1 $ is given by Eq.~(\ref{Usol}) with $\widetilde{\RV}_1 =
\RV_1 $ of Eq.~(\ref{rnlo2}). This is the well-known truncated
analytical NLO solution \cite{FP} which has been employed, for
instance, in refs.~\cite{DFLM,GRV94}.

These two approaches obviously differ in NNLO terms only. The former
procedure introduces more scheme-dependent higher order terms into
the evolution of structure functions like $F_2(x,Q^2)$ or  $g_1(x,Q^2)$
in a general factorization scheme. On the other hand, the latter method
does not solve the evolution equations (\ref{evol1}) literally, but
only in the sense of a power expansion, i.e., up to terms of order
$k \geq 2$. Therefore the first approach may be considered more in the
spirit of the parton model, whereas the second is closer to a
manifestly scheme independent expansion for physical observables.

\subsection{Small-{$x$} resummed evolution}
\label{sect53}

The resummed evolution of the parton distributions includes, to all
orders in $a_s$, the most singular small-$x$ contributions to the
splitting functions $\PV_k$. This inclusion is performed in the orders
beyond the known fixed-order results. Thus the complete  expressions
for $\PV_0(N)$ and $\PV_1(N)$ are used also here, and the difference
to the previous section is restricted to the higher-order matrices
$\PV_{k \geq 2}$. Our notation in this section will directly apply to
the evolution of unpolarized quark and gluon densities. Most of the
subsequent discussion can, however, be easily transferred to the
polarized singlet evolution \cite{BVplb2} by replacing $1/N^{k+l}$ by
$1/(N+1)^{2k+l}$ with correspondingly modified coefficient matrices in
all expansions.

In the present case the most singular small-$x$ terms in the evolution
equations (\ref{evol1}) behave like $ 1/x \: a_s^{k+1} \ln^k x $
\cite{LIPAT} and $1/x \: a_s^{k+1} \ln^{k-1} x $ \cite{CH} as discussed
in Sect.~3. In Mellin-$N$ space these additional resummation
contributions replacing Eq.\ (\ref{PNLO}) read
\beq
\label{Pres2}
 \PV_{k \geq 2}^{\,\rm res}(N) = \frac{\PV_{k}^{\,\rm Lx}}{N^{k+1}}
   + i_{\rm NL} \frac{\PV_{k}^{\,\rm NLx}}{N^k} \: .
\eeq
In particular, the matrix $\PV_{k}^{\,\rm Lx}$ is related to the
expansion coefficients $g_{k,gg}^{(0)}$ in Table 1 by
\beq
 \label{PLx}
  \PV_{k}^{\,\rm Lx} = (4 C_A)^{k+1} g_{k,qq}^{(0)} \left(
  \begin{array}{cc} \! 0       & \! 0 \! \\
                    \! C_F/C_A & \! 1 \! \end{array} \right) \: .
\eeq
$i_{\rm NL}$ indicates whether only these leading small-$x$ pieces are
taken into account (L$x$ resummation), or whether also the next
terms in Eq.~(\ref{Pres2}) are kept (NL$x$ resummation),
\beq
 \label{iNL}
  i_{\rm NL} = \left\{ \begin{array}{l}
  0 \:\:\:\: \mbox{ for {\it Lx } resummation} \\
  1 \:\:\:\: \mbox{ for {\it NLx} resummation} \: . \end{array}\right.
\eeq
Recall that only the upper row of the NL$x$ matrix is completely
known at present \cite{CH}. Results including that part only will be
marked by {\it NLx$_{\, q}$} below.

With respect to the solution of the evolution equations, the situation
is analogous to the fixed-order case, with the expansion parameter
$a_s$ replaced by $N$ at each order $k \geq 2$ of the strong coupling.
The first option is obviously again the direct solution of Eq.\
(\ref{evol2a}), now after inserting Eqs.~(\ref{bNLO}) and (\ref{Pres2}).
To elucidate the generalization of the NLO$_{\rm tr.}$ procedure to the
resummed evolution, consider the splitting functions contribution $
\RV_{k \geq 2} $ (\ref{rmat}) arising from Eq.~(\ref{Pres2}):
\bea
 \label{Rres}
 \RV_{k \geq 2}^{\,\rm res}(N) &=&
   \frac{1}{N^{k+1}}\frac{1}{\beta_0}
   \Big[ \PV_k^{\,\rm Lx} + i_{\rm NL}\, N\, \PV_k^{\,\rm NLx} \Big]
 - \frac{1}{N^k}\frac{\beta_1}{\beta_0^2} \Big[
   \PV_{k-1}^{\,\rm Lx} + i_{\rm NL}\, N\, \PV_{k-1}^{\,\rm NLx} \Big]
 \nonumber\\ & & \mbox{}
 + \frac{1}{N^{k-1}}\frac{1}{\beta_0} \bigg\{ \bigg( \frac{\beta_1}
   {\beta_0} \bigg)^2 - \frac{\beta_2}{\beta_0} \bigg\} \Big[
   \PV_{k-2}^{\,\rm Lx} + i_{\rm NL}\, N\, \PV_{k-2}^{\,\rm NLx} \Big]
 + \ldots \:\:\: .
\eea
The omitted terms involving higher powers of $\beta_1$ and $\beta_2 $,
or $\beta_{k \geq 3}$, are obviously even less singular as the last line
for $N \ra 0 $. Therefore, if the power--counting in $N$ is done on the
level of Eq.~(\ref{evol2a}), one immediately arrives at
\beq
 \label{Rres2}
 \RV_{k \geq 2}^{\,\rm res'}(N) = \frac{1}{\beta_0}\frac{1}{N^{k+1}}
 \bigg[ \PV_k^{\,\rm Lx} + i_{\rm NL}\, N\, \Big( \PV_k^{\,\rm NLx} -
 \frac{\beta_1}{\beta_0} \PV_{k-1}^{\,\rm Lx} \Big) \bigg] \: .
\eeq
Note that in the  NL$x$ (L$x$) case the $\beta$-function
coefficients $\beta_{k \geq 2}$ ($ \beta_{k \geq 1} $) do not contribute
any more, and that $\beta_1 $ occurs only linearly in the former case.
Thus $\PV_1$, which does not exhibit an L$x$ contribution, does not
enter Eq.~(\ref{Rres2}) in the present unpolarized case. All this is
completely analogous to the $\RV$ matrices (\ref{rnlo2}) for the
NLO$_{\rm tr.}$ evolution.

Before we turn to the $\UV $ matrix for this second procedure, it is
instructive to consider a small-$x$ approximation to the unpolarized
L$x$ evolution in this approach. Unlike in any other QCD singlet
case, including the polarized leading small-$x$ resummation \cite
{BVplb2}, the splitting function combinations $\RV_{k,k' \geq 2}$ do
commute here: $[\RV_k(N), \RV_{k'}(N)] = 0$, due to the simple
structure of the matrix (\ref{PLx}). This is still not sufficient for a
closed solution of the evolution equation (\ref{evol2a}), unless one
also keeps the leading small-$x$ contributions to $\PV_{0,1}(N)$ only.
Then, however, one eigenvalue of $\PV_0$ vanishes, resulting in
\beq
  \PV_{0}^{\, x \ra 0} = \frac{4 C_A}{N}
  \left( \begin{array}{cc} \! 0       & \! 0 \! \\
          	           \! C_F/C_A & \! 1 \! \end{array}\right)
  = \frac{4 C_A}{N}\, \eV_{+}^{\, x \ra 0} \: .
\eeq
Using Eqs.~(\ref{GAML}) one obtains with this additional approximation
\bea
 \qV_{\rm Lx}^{\,\rm approx.}(a_s,N) &\, =\, & \exp\bigg[
  \frac{1}{2\beta_0} \int_{a_0}^{a_s} \! da \, \frac{1}{a^2}
  \GGV_L (a,N) \bigg] \qV (a_0, N)  \nonumber \\
 &\, =\, & \exp \left(\frac{12 L}{\beta_0 N}\right) \left [ 1 +
  \sum_{l=1}^{\infty} \frac{d_l(a_s,a_0)}{N^l}\right]
  \eV_+^{\, x \ra 0} \,\qV (a_0, N) \: ,
\label{qaprN}
\eea
analogous to the complete resummed non-singlet solution, see refs.\
\cite{BVplb1}. This simple approximate expression, however, does
of course not lead to any quark evolution. Eq.~(\ref{qaprN}) can be
completely transformed to $x$-space, cf.~ref.~\cite{JBKT},
\beq
  \qV_{\rm Lx}^{\,\rm approx.} (a_s,x) =
  F(a_s,a_0,x) \otimes  \eV_+^{\, x \ra 0} \,\qV(a_0,x) \: ,
\label{qaprx}
\eeq
with
\beq
F(a_s,a_0,x) = \frac{1}{x} \bigg [ \delta(1-x) +
\sqrt{\frac{12 L}{\beta_0 \log(1/x)}}I_1(z)
+ \sum_{l=1}^{\infty} d_l(a_s,a_0) \left(\frac{\beta_0 \log(1/x)}{12 L}
\right)^{(l-1)/2} I_{l-1}(z) \bigg],
\label{eqfa}
\eeq
where
\beq
z = 2 \left [\frac{12 L}{\beta_0} \log \left(\frac{1}{x}\right)
\right]^{1/2}~,~~~~~L = \log \left(\frac{a_s}{a_0}\right)~,
\eeq
and $I_{\nu}(z)$ denotes the Bessel functions of imaginary argument.
Similar expressions, e.g.\ in the double-logarithmic approximations,
have been studied in detail long ago~\cite{DLFO} and were also
considered recently~\cite{BF0}. As compared to the complete L$x$
solution, however, the approximation   (\ref{qaprx}) yields gluon
densities which are typically too large by a factor of 2 for an
evolution from 4 to 100 GeV$^2$. Therefore we will not apply this
approach in the following.

We now proceed with the general resummed solution corresponding to the
truncated NLO treatment where $\UV_{k \geq 2} = 0$, see Eq.~(\ref
{sol2}). The generalization to the present case is to keep only those
terms of $\UV_{k \geq 2}$, which arise from the L$x$ and NL$x$ pieces
of $\RV_1$ and of $\RV_{k \geq 2}$ in Eq.~(\ref{Rres}). Hence the NLO
coefficient $\UV '_1 = \UV _1$ is supplemented by
\beq
 \big[ \UV'_2, \RV_0 \big] = \RV_{2}^{\,\rm res'} + \RV_1^S \UV_1^S
 + 2 \UV'_2
\eeq
etc. Here $\RV_1^S$ denotes the small-$x$ contribution of $\RV_1$
\beq
 \label{Rres1}
 \RV_1^S(N) = \frac{1}{\beta_0}\frac{1}{N^2}
 \bigg[ \PV_1^{\,\rm Lx} + i_{\rm NL}\, N\, \Big( \PV_1^{\,\rm NLx} -
 \frac{\beta_1}{\beta_0} \PV_0^{\,\rm Lx} \Big) \bigg] \: ,
\eeq
and the corresponding expansion coefficient $\UV_1^S$ is given by
\beq
 \big[ \UV_1^S, \RV_0 \big] = \RV_1^S + \UV_1^S \: .
\eeq
The final step analogous to the NLO$_{\rm tr.}$ method is to keep the
non-(N)L$x$ parts of $\UV_1$ only linearly also in the inverse matrix
$\UV^{-1}$ of Eq.~(\ref{sol1}). This leads to
\bea
\label{uexp}
 \UV^{\prime\: -1}(a,N) &=&
  \Big[ 1 + a\,\UV_1^S(N) + \sum_{k=2}^{\infty} a^k \UV'_k (N)
  \Big]^{-1} + a\, \{ \UV_1^S(N) - \UV'_1(N) \} \nonumber \\
 &\equiv & \UV_S^{-1}(a,N) - \{ 1 - a \UV_1^S(N)\} + 1 - a\,\UV_1(N)
  \: .
\eea
The last two terms represent the truncated NLO contribution. Insertion
of this decomposition into Eq.~(\ref{sol1}) finally yields (with $\LV
\equiv \LV (a_s,a_0,N)$ for brevity)
\bea
\label{sol3}
 \qV^{\,\rm res}_{\,\rm tr.} (a_s,N) &\! =\, & \Big[ \LV + a_s \UV_1(N)
 \LV - a_0 \LV \UV_1(N) \Big] \: \qV (a_0,N) \\
 &\, +\! & \Big[ \UV_S (a_s,N) \LV \UV_S^{-1}(a_0,N) -
 \LV - a_s \UV_1^S(N) \LV + a_0 \LV \UV_1^S(N) \Big] \: \qV (a_0,N)
 \: . \nonumber
\eea
The first line is the NLO$_{\rm tr.}$ result (\ref{sol2}), the second
line represents the resummation correction.

\subsection{The photonic solution}
\label{sect54}

We now turn to the parton distributions of the photon. In terms of the
running coupling, the corresponding inhomogeneous evolution equation
(\ref{evolx}) reads
\bea
\label{phev2}
 \frac{\partial \qV^{\gamma}(a_s,N)}{\partial a_s} &=& \frac{a_{\rm em}
  \{ \kV_0(N) + a_s \kV_1(N) + a_s^2 \kV_2(N)+ \ldots \} }
  {-a_s^2 \,\beta_0 - a_s^3 \,\beta_1 - a_s^4 \,\beta_2 \ldots} +
  \mbox{ had.} \nonumber \\
 &=& -\frac{a_{\rm em}}{a_s^2} \bigg[ \KV_0 (N) + \sum_{l=1}^{\infty} a_s^l
  \KV_l (N) \bigg] + \mbox{ had. } \:\equiv\: \KV(a_s,N) +
  \mbox{ had.} \:\: .
\eea
Here $a_{\rm em} = \alpha /4\pi $ denotes the electromagnetic fine
structure constant, and analogously to Eqs.~(\ref{r0}) and (\ref{rmat})
we have introduced
\bea
\label{k0}
  \KV_0 &\equiv & \frac{1}{\beta_0} \kV_0 \: , \\
  \KV_l &\equiv & \frac{1}{\beta_0} \kV_l - \sum_{i=1}^{l}
                  \frac{\beta_{i}} {\beta_0} \KV_{l-i} \: .
\label{kvec}
\eea
Finally 'had.' stands for the r.h.s.\ of the hadronic evolution equation
(\ref{evol2a}), with $\qV $ replaced by $\qV^{\gamma}$. The homogeneous
component, $\qV_{\rm hom.}$, of the solution of Eq.~(\ref{phev2}) is as
derived in Sect.~5.1 -- 5.3. Hence only the inhomogeneous part,
$\qV_{\rm inh.} = \qV^{\gamma} - \qV_{\rm hom.}$ with $\qV_{\rm inh.}
(a_0,N) = 0$, needs to be discussed here. This solution can be
represented in terms of the hadronic evolution operator (\ref{sol1}) as
\beq
\label{phs1}
  \qV_{\rm inh.}(a_s,N) = \UV(a_s,N)\, a_s^{{\footnotesize \, -\RV}_0
  (N)} \int_{a_0}^{a_s} \! da \, a^{{\footnotesize \,\RV}_0 (N)}
  \UV^{-1} (a,N) \KV(a,N) \: .
\eeq

For the iterative solutions the remaining integral can be performed
numerically. In the truncated procedures, on the other hand,
$\UV^{-1}(a,N)$ has been expanded in Eqs.~(\ref{sol2}) and (\ref{uexp}).
In these cases the NLO photonic splitting functions, $k_1^{\, i} \equiv
P_1^{\, i\gamma}$, should be treated in the same way as their hadronic
counterparts $ P_1^{\, ij}$ previously, reducing Eq.~(\ref{kvec}) to
\beq
\label{knlo2}
 \KV_1^{\prime} = \KV_1 = \frac{1}{\beta_0} \Big( \kV_1 -
 \frac{\beta_{1}} {\beta_0} \kV_{0} \Big) \: , \:\:\:\:\:
 \KV_{l \geq 2}^{\prime} = 0 \: .
\eeq
In the following we will confine ourselves to physical factorization
schemes like the DIS$_{\gamma}$ scheme \cite{GRVph1} or the DIS
scheme, where the photonic coefficient function $C_{2,\gamma}$ has been
absorbed into the quark distributions. In these schemes $\KV (a_s,N)$
does not receive any L$x$ and NL$x$ resummation corrections beyond the
leading order, as discussed in Sect.~3. Thus Eq.\ (\ref{knlo2}) applies
to the NLO$_{\rm tr.}$ photon evolution as well as to the corresponding
resummed case.

Inserting $\UV^{-1}_{\rm NLO'}(a_s,N) = 1 - a_s \UV_1(N)$ and the
expansion (\ref{knlo2}) into the inhomogeneous solution (\ref{phs1}),
the $a_s$ integration becomes obvious and one arrives at \cite{GRVph1}
\bea
\label{phs2}
 \frac{1}{a_{\rm em}} \qV_{\rm inh.}^{\,\rm NLO'}(a_s,N) &=&
 \frac{1}{a_s} [1 + a_s \UV_1(N)]\, (1 - \frac{a_s}{a_0} \LV )\,
 [1-\RV_0(N)]^{-1} \KV_0(N) \nonumber \\
 & & \mbox{} - (1-\LV)\, \RV_0^{-1}(N)\, [\KV_1(N)- \UV_1(N) \KV_0(N)]
 \, + \, O(a_s) \: .
\eea
In the resummed case a numerical integration remains over the all-order
part of $\UV^{\prime\: -1}$ in Eq.~(\ref{uexp}). Defining
\beq
 \Delta_{\rm res}(a_s,a_0,N) = - \int_{a_0}^{a_s} \! \frac{da}{a^2}
 \LV^{-1}(a,a_0,N) \Big( \UV_S^{-1}(a,N) - 1 + a \UV_1^S(N) \Big)
 \KV_0(N)
\eeq
the solution is, again using $\LV \equiv \LV (a_s,a_0,N)$, given by
\bea
\label{phs3}
 \lefteqn{ \frac{1}{a_{\rm em}} \qV_{\rm inh.}^{\,\rm res'}(a_s,N)
  \: = \: \frac{1}{a_e} \qV_{\rm inh.}^{\,\rm NLO'}(a_s,N) +
  \UV_s(a_s,N) \,\Delta_{\rm res}(a_s,a_0,N) } \\
 & & \mbox{}+ \UV_s(a_s,N) \left[ \frac{1}{a_s}(1 -\frac{a_s}{a_0}\LV )
  \, [1-\RV_0(N)]^{-1}\,  \KV_0(N) \, +\, (1-\LV)\, \RV_0^{-1}(N)\,
  \UV_1^S(N)\, \KV_0(N) \right] \nonumber \\
 & & \mbox{}- [1 + a_s \UV_1^S(N)]\, \frac{1}{a_s}\, (1 -\frac{a_s}{a_0}
  \LV )\, [1-\RV_0(N)]^{-1}\, \KV_0(N)\, -\, (1-\LV)\, \RV_0^{-1}(N)\,
  \UV_1^S(N)\, \KV_0(N) \nonumber \: .
\eea
This relation completes the Mellin-$N$ solutions of the fixed-order
and resummed, hadronic and photonic evolution equations. We are now
prepared to investigate the quantitative impact of the various
approximations, for both the splitting functions and the solutions,
on the parton densities and structure functions.
\section{Numerical results}
\label{sect6}
\setcounter{equation}{0}

\noindent
In the following we study the numerical consequences of the fixed-order
and resummed evolution kernels on the evolution of structure functions
and some aspects related to potential uncertainties. Despite the
impressive amount of small-$x$ structure function data already
collected at HERA \cite{HEXP,FLEXP}, the present investigation does
not aim at a comparative data analysis. Such an effort would require
quite some flexibility of the non-perturbative initial distributions,
especially for the gluon density which is only rather indirectly
constrained by measurements of $F_2$ and $F_L$. A detailed data
analysis requires rather precise and independent constraints on the
small-$x$ behavior of the gluon density, which are not yet provided by
current measurements at HERA. A thorough implementation of heavy flavor
(charm) mass effects in the resummation framework would be required as
well. These mass effects are non-negligible at small $x$, where the
charm contribution to $F_2$ and $F_L$ is substantial, in spite of the
very large hadronic invariant mass, $W^2 \gg 4 m_c^2$, cf.~refs.\
\cite{charm}. Both of these issues lie beyond the scope of the present
paper. Since some of the resummation corrections turn out to be very
large one would like to know as well the next-order resummed
corrections to perform a detailed data analysis.

In the following, therefore, the impact of the various anomalous
dimensions and Wilson coefficients is instead illustrated for fixed
initial parton densities of the proton and the photon. Accordingly all
calculations are performed using the same values for $\alpha_s(Q^2)$.
Specifically, the NLO relation (\ref{alps}) is employed with $\Lambda_
{N_f = 4} = 0.23$ GeV above $Q^2 = m_c^2 = (1.5 \mbox{ GeV})^2$ and,
by continuity of $\alpha_s (Q^2)$, with $\Lambda_{N_f = 3} = 0.30$
GeV below that scale. Above (below) $Q^2 = m_c^2$ the evolution
equations are solved for four (three) massless flavors, respectively,
with $c(x,m_c^2) = \bar{c}(x,m_c^2) = 0$. The small effects of the
bottom flavor are entirely neglected. All subsequent results are
derived in the DIS scheme discussed above, with the truncated solutions
of Sect.~5 chosen as default. Only the singlet resummations described
in Sect.~3 are taken into account, since the non-singlet contributions
are suppressed at small $x$ in the present unpolarized case, see
Fig.~5, and its resummation correction is very small, cf.~Sect.~2.2.

\subsection{Proton structure: fixed-order evolution}
\label{sect61}

\noindent
Let us first consider the leading and next-to-leading order evolution
of hadronic parton densities, putting emphasis on the small-$x$ region.
As the value of $\Lambda_{N_f = 4}$ given above, the initial
distributions for our proton studies are adopted from the MRS(A$'$)
global fit \cite{MRSA} at a reference scale $Q_0^2= 4 \mbox{ GeV}^2$.
For the present purpose, the most relevant feature of these input
densities is their small-$x$ behavior which has been constrained by
previous HERA data:
\bea
  xg^p(x, Q_0^2) \,\propto\, x\Sigma^p (x, Q_0^2) \,\propto\, x^{-0.17}
  \:\:\: \mbox{ for } \:\: x \rightarrow 0 \: .
\eea
Recall that, unlike the gluon distribution, the DIS-scheme quark
densities represent observables.

The LO and NLO small-$x$ evolution of $x\Sigma^p$ and $xg^{p}$ to
$Q^2 = 10 $ and 100 GeV$^2$ is shown in Fig.~5 together with the
initial distributions. The LO curves have been calculated, as indicated
above, using the NLO input densities and $\alpha_s$ values in
Eq.~(\ref{sol0}). Hence they do not represent results of an independent
leading-order analysis, but directly illustrate the importance of the
NLO terms relative to the LO contribution in Eq.~(\ref{sol2}). One
notices that the perturbative stability of the presently available
fixed-order evolution is theoretically satisfactory also at very low
values of~$x$. For instance, the NLO/LO ratio amounts to 1.25 (0.87)
for the singlet (gluon) density at $Q^2 = 100 \mbox{ GeV}^2 $ and
$ x = 10^{-4}$. Furthermore the numerical differences between the
expanded solution (\ref{sol2}) and the iterative approach (\ref{rnlo1})
to the NLO evolution equations can be considered as absolute lower
limits on the uncertainties due to the unknown higher-order splitting
functions. These offsets reach 3\% at $x \simeq 10^{-5}$, while
amounting to less than 1\% for $x \geq 10^{-3}$, see also Fig.~6 below%
\footnote{In previous comparisons \cite{RH2,Progs2} deviations of up
 to 8\% were found between these solutions. These large effects
 originated in the unfortunate choice of a traditional approximate NLO
 expression for $\alpha_s(Q^2)$, showing that the representation of
 the NLO solution of Eq.~(\ref{eval}) tends to be relevant at the
 current accuracy of the data.}.
Thus one may roughly expect a $ 5-10\% $ small-$x$ effect from the
3-loop anomalous dimensions, if fixed-order  renormalization group
improved  perturbation theory remains the appropriate framework down to
$x \gsim 10^{-5}$. Such an estimate is also corroborated by studies of
the factorization scale dependence of $F_2$ at small $x$~\cite{RH2}.

It is conceivable, however, that the NLO contributions to the small-$x$
anomalous dimensions are untypically small (as, for instance, $1/N^2$
terms are absent in $\gamma_1$, cf.~Eqs.~(4.2) and Table~1). In this
context it is instructive to study the convergence of (formal)
small-$x$ expansions of anomalous dimensions and Wilson coefficients
into the series
\beq
 \varphi(a_s, N) = \sum_{l=1}^{\infty} a_s^l \left[ \frac
 {\varphi_l^{\,\rm Lx}}{N^l} + \frac{\varphi_l^{\,\rm NLx}}{N^{l-1}}
 + \frac{\varphi_l^{\,\rm NNLx}}{N^{l-2}} + \ldots \right]
\label{eqAPP}
\eeq
already at the LO and NLO level, where the full results are available.
Also shown in Fig.~5, therefore, is the NL$x_q$ approximation to the
leading-order evolution, for which just the $1/N$ terms of $\gamma
_{gg}^{\rm LO}$ and $\gamma_{gq}^{\rm LO}$ are kept together with the
leading $N \!\rightarrow\! 0$ constants in $\gamma_{qg}^{\rm LO}$ and
$\gamma_{qq}^{\rm LO}$, see Eqs.~(4.1). Note that this procedure is
close to the well-known double-logarithmic approximation, cf. Sect.~5.3.
As can be seen from the figure, this first approximation is very poor:
$x\Sigma$ ($xg$) exceed the full LO results by factors of about 1.7
(2.2), respectively, rather uniformly in $x$ at $Q^2 = 100 \mbox{ GeV}
^2$, without any appreciable sign of improvement for decreasing values
of~$x$.

Hence the question arises how many terms in the small-$x$ expansion
are required for arriving at a reasonably accurate representation of
the complete fixed-order results. Accordingly, Fig.~6 displays the
ratios $\Sigma^{\rm approx.}/ \Sigma^{\rm full}$ and $g^{\rm approx.}/
g^{\rm full}$ for the LO and NLO evolutions with an increasing number
of terms taken into account in the expansion (\ref{eqAPP}) of
$\gamma_0$ and $\gamma_1$ (in NLO the complete expression for
$\gamma_0$ has been employed for all curves). One finds that in general
three to four non-trivial small-$x$ terms, i.e., contributions up to
N$^2$L$x$ at LO and N$^3$L$x$ at NLO, are needed to achieve an accuracy
of better than 10\%. The NLO situation is not a peculiarity of the DIS
scheme chosen here, as a corresponding \MSbar\ analysis using
Eqs.~(4.3) yields similar results. Note that an interesting pattern
emerges in both fixed-order cases: the approximate results alternate
around the exact values with decreasing amplitude. If such a pattern
were to persist to higher orders in $\alpha_s$, a first reliable
estimate of their possible impact could be derived once two more
non-trivial terms in all small-$x$ expansions were known. This aspect
may be of relevance for the resummed evolution addressed in the
following.

\subsection{Resummed evolution}
\label{sect62}

We now turn to the effects and the relative importance of the L$x$
\cite{LIPAT} and
NL$x$~[8--10]
higher-order contributions to
the splitting functions discussed in Sect.~3. In the present subsection
the momentum sum rule (\ref{conserv}) is restored by prescription (A)
of Eq.~(4.4), i.e., $P^{gg}_{k}$ and $P^{qq}_{k}$ are supplemented by
appropriate $\delta (1-x)$ terms at all orders $k \geq 2$. This
procedure is the one with the least impact on the small-$x$ results.
Without any subtraction the sum rule would be violated by about 1\% and
6\% at $Q^2 = 100 \mbox{ GeV}^2$ for the L$x$ and NL$x_q$ resummed
evolutions, respectively, of our MRS(A$'$) initial distributions.

The resulting evolutions of the singlet quark and gluon densities are
compared to the NLO distributions in Fig.~7 for $Q^2 = 10 $ and $100
\mbox{ GeV}^2$. The relative importance of the available gluonic (lower
row) anomalous dimensions is illustrated in Fig.~8(a). Consider first
the effect of the L$x$ corrections \cite{LIPAT}. These terms exert an
appreciable influence on the gluon evolution, but much less on the
quark densities in the kinematic region covered by the figure. At $x =
10^{-4}$ and $Q^2 = 100 \mbox{ GeV}^2$, e.g., ratios of $g^{\rm Lx} /
g^{\rm NLO} = 1.24 $ and $\Sigma^{\rm Lx} / \Sigma^{\rm NLO} = 1.07 $
are obtained. This pattern obviously arises from the matrix structure
of the L$x$ kernel (\ref{GAML}); only at higher scales the quark effect
fully approaches the gluon enhancement.

The inclusion of the NL$x_q$ terms \cite{CH}, i.e.\ the upper row
entries in Eq.~(\ref{GAMNL}), leads to only a small additional effect
on $xg(x,Q^2)$. The impact of these terms on $x\Sigma(x,Q^2)$ is,
however, exceedingly large, as already evident from Fig.~2. These
effects have been illustrated before, cf.~refs.~\cite{EHW,BRV}, partly
using different parametrizations for the input distributions at the
starting scale $Q^2_0$. The resulting enhancement with respect to the
NLO evolution amounts to a factor of 2.8,
for example, at $x = 10^{-4}$ and $Q^2 = 100 \mbox{ GeV}^2$. This huge
correction is indeed entirely driven by the quarkonic anomalous
dimensions, as also illustrated in Fig.~8(a): any `reasonable' change
of the gluonic splitting functions affects $x\Sigma$ by at most about
10\%. Since only this one resummation contribution is known for the
dominant upper-row quantities, a theory based estimate along the lines
of Sect.~6.1 is not yet possible for the resummed quark distributions,
and hence for the most important structure function, $F_2$. We will
therefore resort to the sum-rule prescriptions of Sect.~4 in the next
subsection.

In the gluonic sector, on the other hand, the theoretical situation has
been improved recently by refs.~\cite{CC1,CC2}, see Sect.~3. We remind
the reader that the latter findings for $\gamma_{gg}$, although
indicative, are not final yet, since the so-called energy-scale
dependent NL$x$ terms have still to be calculated.
The effects of the known next-to-leading contributions are
also presented in Figs.~7 and 8(a). The well-established $q\bar{q}$
contribution to $\gamma_{gg}^{(1)}$ \cite{CC1}, which is not expected
to yield the largest subdominant terms, already removes more
than half of the L$x$ effects on the gluon density at $x \lsim 10^{-4}$
for $Q^2 = 100 \mbox{ GeV}^2$, see Fig.~8(a). The energy-scale
independent gluonic contribution \cite{CC2} overcompensates the
enhancement by the L$x$ and NL$x_q$ terms at all $x$ in Fig.~7. As
expected from Fig.~3, these terms are that large that they even cause a
sign change in the slope of the gluon evolution for $x \lsim 10^{-4}$.
It seems natural to expect that yet missing terms either in NL$x$ or
unknown terms emerging in higher orders correct this behavior again.
Thus a first uncertainty band of the possible resummation effects on
$xg(x,Q^2)$ seems close to completion.
In this context it should be recalled that the NL$x$ anomalous
dimension matrix is not yet complete, as $\gamma_{gq}^{(1)}$ still
remains uncalculated. Note, however, that $\gamma_{gq}^{(0)}$ has an
impact of less than 10\% on both $x\Sigma$ and $xg$ in the L$x$ and
NL$x_q$ evolutions, cf.~Fig.~8(a). Hence $\gamma_{gq}^{(1)}$ is
presumably not a major source of uncertainty, as one may expect a
rather moderate effect of this quantity as well$\,$%
\footnote{This expectation is also supported by the fact that
 $\gamma_{gq}^{(1)}$ does not contribute to the eigenvalues of the
 resummed anomalous dimension matrix up to the NL$x$ level, see
 Eq.~(\ref{eqapp}).}.

As in the NLO case of Sect.~6.1, the differences between the iterative
and the truncated solutions, Eqs.~(\ref{Rres}) and (\ref{sol3}), of the
evolution equations should yield a lower limit on the uncertainty due
to missing terms in the anomalous dimensions. In fact, if the present
small-$x$ resummations collected the most relevant higher-order terms,
a reduction of these offsets should take place with respect to the NLO
evolution. The corresponding results are depicted in Fig.~8(b). While
staying on the same level as in the NLO case for the L$x$ evolution,
the offsets increase significantly as soon as the NL$x$ terms are
included, in particular for the singlet quark density: ratios $\Sigma
^{\rm iter'd} / \Sigma^{\rm trunc'd}$ of up to about 10\% are found.
This decreased stability may point to a larger uncertainty of the huge
NL$x$ quark enhancement.

\subsection{Structure functions and less singular terms}
\label{sect63}

\vspace{1mm}
\noindent
Less singular (subleading) contributions to the anomalous dimensions,
i.e., terms which do not exhibit the leading $N \!\rightarrow\! 0$
behavior, are vitally important for the LO and NLO evolution at small
$x$, as demonstrated in Fig.~6: three to four terms in the expansion
(\ref{eqAPP}) are required for a good representation. In higher orders
of $\alpha_s$ the leading $N \!\rightarrow\! 0 $ poles become more
singular, but so do the subleading contributions, and the number of
singular pieces increases. There is, therefore, no obvious reason to
expect terms less singular in $N$ to be unimportant at low $x$ in
all-order approaches.
At the present stage of the theoretical development, however, one has
to rely on reasonable estimates for obtaining  a first impression
of their possible impact. For this purpose, we employ the
momentum sum-rule prescriptions (C) and (D) of Eq.~(4.4) for all
anomalous dimensions with only one all-order term known presently,
\beq
\label{eq63}
  \gamma^{\, ij}_{k \geq 2}(N) \rightarrow \gamma^{\, ij}_{k \geq 2}(N)
  (1 -2N + N^{\alpha}) \:\:\: \mbox{ for }\:\: ij = qq,\, qg,\, gq\, .
\eeq
Here $\alpha$ = 2[3] for the prescriptions (C)[(D)]. In $\gamma_{gg}$
we adopt the presently known NL$x$ contributions, which are taken from
Table~1. Hence only the $N^2$ or $N^3$ terms in this quantity are
adjusted according to Eq.~(\ref{conserv}). In view of the structure of
the $N$-expansions of the LO and NLO terms estimates like
Eq.~(\ref{eq63}) are conservative, i.e., they might underestimate the
present uncertainties.

In Figure 9 the resulting singlet quark and gluon densities are
compared at NL$x$ accuracy to distributions evolved with prescription
(A).  The  subleading terms of the ansatz (D) are
sufficient to overcompensate the huge leading resummation effect on
$x\Sigma(x,Q^2)$ slightly. E.g., the NL$x^ {(D)}$ result falls about
10\% short of the NLO distribution at $x = 10^{-4}$ and $Q^2 = 100
\mbox{ GeV}^2$. Note that even the difference between the prescriptions
(C) and (D), arising from the replacement of parametrically small
N$^3$L$x$ by N$^4$L$x$ terms in the quarkonic anomalous dimensions,
proves rather appreciable. This situation is similar for $xg(xQ^2)$,
where the effects of the sum-rule induced terms are positive because of
the very large negative $\gamma_{gg}^{(1)}$ entries, cf.~Table~1. The
order of the curves is different here as compared to $x\Sigma$, since
$\gamma_{gg}$ differs between the cases (A), (C), and (D) only in the
third term of the small-$x$ expansion, unlike $\gamma_{qg}$ which
dominates the quark evolution. Although definite conclusions cannot be
drawn from these prescription-dependent results, they nevertheless
indicate clearly that the $1/N$ expansion (\ref{eqAPP}) behaves similar
as in the fixed-order cases.

We now turn to the proton structure functions $F_2$ and $F_L$. Their
small-$x$ behavior, as obtained from the parton densities just
discussed, is displayed in Figure 10. Since our calculations are
performed in the DIS scheme, $F_2$ is very closely related to the quark
singlet distribution at small $x$. Thus the left side of Fig.~10
exhibits a pattern very similar to Fig.~9. The longitudinal structure
function, on the other hand, in addition involves the resummed
coefficient functions
\beq
\label{eq64}
 C_L(\alpha_s,N) = a_s C_{L,0}(N) + a_s^2 C_{L,1}(N) + \frac{4}{3} N_f
 \, a_s \sum_{k=2}^{\infty} c_k^L \left( \frac{\overline{\alpha}_s}{N}
 \right)^k \: .
\eeq
$C_{L,0}$ and $C_{L,1}$ represent the leading and next-to-leading
order~\cite{CL1,CO2} coefficient functions, and the gluon and pure
singlet resummation coefficients \cite{CH} $c_k^{\, L}$ are given in
Table~1 and Eq.~(\ref{eqS5}).

The additional corrections due to the coefficients $c_k^L$ are,
in fact, very large at the lower $Q^2$ values shown: even the
cross-section positivity constraint $F_L < F_2$ is violated for $x
\lsim 3 \cdot 10^{-4}$ at $Q^2 \simeq 4 \mbox{ GeV}^2$ for the
MRS(A$'$) initial distributions. At high $Q^2 \gsim 100 \mbox{ GeV}^2$
the effects of the coefficient functions and of the parton evolution
become comparable due to the decrease of $\alpha_s$ in $C_L(\alpha_s,
N)$. The size of the low-$Q^2$ effect shown in Fig.~10 (upper dotted
and dashed-dotted curves), however, requires sizeable corrections by
yet unknown higher-order terms in the small-$x$ resummation of $C_L$
or a large adjustment of the input gluon density. In fact, also the
coefficient functions can be expected to receive relevant subleading
corrections which are unknown at present. In order to derive a first
estimate on their possible impact, the $F_L$ calculations have been
repeated with
\beq
  c_{k\geq 2}^{\, L} \rightarrow c_{k\geq 2}^{\, L} \, (1 -2N) \: .
\eeq
The results of these calculations are shown in Fig.~10 (lower curves).
This moderate correction term leads to an even drastic overcompensation
of the leading resummation effect at low $Q^2$. This shows that for a
more detailed understanding of the small-$x$ behavior of $F_L(x,Q^2)$
the next-order small-$x$ resummed corrections are required.
On the other hand, direct measurements of $F_L(x,Q^2)\,$%
\footnote{The measurements of $F_L(x,Q^2)$, ref.~\cite{FLEXP}, are
 `indirect' and correlated with the $F_2$ measurement. Their present
 experimental errors are still large. More precise results are expected
 from the data of the 1997 HERA run.}
by the HERA experiments could help to constrain the size of missing
terms in the coefficient functions.

\subsection{Photon structure at small-{\boldmath $x$}}
\label{sect64}

\noindent
We now address, finally, the small-$x$ evolution of the parton
densities of the real photon. As outlined in Sect.~2, this evolution
includes a specific inhomogeneous (`pointlike') piece in addition to
the homogeneous (`hadronic') component. Whereas the latter behaves
rather similar to the proton's parton distribution considered in the
preceding subsections, the former is completely calculable in
perturbation theory up to its dependence on the starting scale $Q_0^2$.
As discussed in Sect.~3 one may study the evolution of the photon
structure function in a DIS scheme, where the inhomogeneous part does
not involve any new resummed splitting functions at the present level of
accuracy.
It does, however, probe the resummed hadronic evolution matrix in a
specific, different manner, cf.~Sect.~5, and thus provides an
additional laboratory for studying the possible effects of the
small-$x$ resummations.

The reference scale $Q_0^2$ takes a somewhat different character in the
photon case than in the  pure hadronic evolution. It is a free parameter
for the solution of the evolution equations, still, but only for certain
choices of $Q_0^2$ can the separation between the homogeneous and
inhomogeneous pieces approximately reflect the physical decomposition
into the non-perturbative component, induced, e.g., by vector meson
dominance (VMD), and a perturbative contribution. In fact this physical
decomposition leads to $Q_0^2 < 1 \mbox{ GeV}^2$ \cite{GRVph2,OTH1},
for a recent overview cf. ref.~\cite{avEgm}. We therefore
choose $Q_0^2 = 1 \mbox{ GeV}^2$ in the following, unlike the proton
case of Sect.~6.1--6.3. At this scale we adopt the NLO photonic parton
distributions of GRV \cite{GRVph2}, as this is the only available NLO
set with a HERA-like small-$x$ rise of the hadronic component. The
low-$x$ behavior  of these singlet and gluon densities is not given by
a simple power law (6.1), however, but can approximately be written as
\bea
  \Sigma^{\gamma} (x, Q_0^2) \,\sim\, x^{-0.22} \:\: ,\:\:
  g^{\gamma} (x, Q_0^2) \,\sim \, x^{-0.13 \ldots -0.22}
  \:\:\: \mbox{ for } \:\: 10^{-4} \lsim x < 10^{-2} \: .
\eea
Here the effective rising power of $xg^{\gamma}$ decreases with
decreasing $x$, cf.~ref.~\cite{GRV94}.

The fixed-order evolution of $x\Sigma^{\gamma}$ and $xg^{\gamma}$
is recalled in Fig.~11. As in the proton case, the LO solution has been
calculated using the NLO (DIS scheme) initial distributions and the NLO
values for $\alpha_s(Q^2)$. The NLO/LO difference is slightly larger
than in Fig.~5 due to the larger values of the coupling constant
involved. Also shown in the figure is the NLO hadronic VMD contribution
which is suppressed (in particular in the quark case) at large $x$, but
dominant in the small-$x$ regime: it still amounts to about 80\% of the
full results for $x < 10^{-3}$ at $Q^2 = 100 \mbox{ GeV}^2$. Therefore
one may expect a similar rise of $F_2^{\gamma}$ as observed for $F_2^p$
at HERA \cite{HEXP}. We shall consider now how the resummation
corrections affect this picture.

Fig.~12 present the effects of the various resummed small-$x$ terms
[4,8--10]
on the evolution of the singlet quark and gluon
distributions. The full results and the inhomogeneous contributions for
$Q_0^2 = 1 \mbox{ GeV}^2$ are separately shown. These effects are
considerably larger than those in Fig.~7: the NL$x$/NLO ratios reach
factors of about 8 and 2 here for $x\Sigma^{\gamma}$ and $xg^{\gamma}$,
respectively, at $x = 10^{-4}$ and $Q^2 = 100 \mbox{ GeV}^2$. The
dominant source of this greater enhancement are again the larger
$\alpha_s$ values implied by the lower choice of $Q_0^2$. The
inhomogeneous components are still suppressed, although they are even
more affected by the resummation corrections, as factors of up to 15
and 8 are found for the NL$x$/NLO ratios of $\Sigma_{\rm inhom.}$ and
$g_{\,\rm inhom.}$, respectively$\,$%
\footnote{In contrast, the NLO/LO ratio  not shown in the figure
is on a rather normal level.}.
Recall that these latter results do not depend on any non-perturbative
input distributions. Note also that $xg_{\,\rm inhom.}$ is much less
affected by the $\gamma_{gg}^{(1)}$ corrections \cite{CC1,CC2}, since
the main `driving term' of the inhomogeneous solution is the purely
quarkonic quantity $\kV$, cf.~Eq.~(\ref{phs3}).

The possible effects of less singular terms, using the same momentum
sum-rule prescription as in the proton evolution, are illustrated in
Fig.~13 for $F_2^{\gamma}$ and $xg^{\gamma}$. The general pattern for
the total results is analogous to the purely hadronic case of Figs.~9
and 10. The relevance of subleading corrections, however, is even more
enhanced than that of the leading terms: $F_2^{\gamma}$ falls far below
the NLO calculation for the ansatz (D), and the breakdown of the gluon
evolution in NL$x^{(A)}$ already takes place at $x \lsim 10^{-3}$. On
the other hand, the less singular terms are much less effective in
$F_{2,\rm inhom.}$ at small $x$. In hadron-like cases their importance
is magnified by the convolution with the (soft) parton densities, cf.\
Figs.~2 and 3. Here, however, the function $xk_0^q \propto x [1 +
(1-x)^2]$, which plays the role of an `input distribution', is very
hard. Hence $F_{2,\rm inhom.}$ comes closer to a local probe of the
small-$x$ splitting functions than any inclusive hadronic quantity.

Nevertheless the inhomogeneous part remains much smaller than than
homogeneous piece of $F_2^{\gamma}$ for most scenarios  of Fig.~13.
It should be noted, however, that the latter contribution may be
suppressed down to about the NLO results by a different choice of
$xg^{\gamma}$. It is conceivable, therefore, that $F_{2,\rm inhom.}$ is
much more important in the resummed evolution than in the fixed-order
case discussed above. Indeed, a small-$x$ $F_2^{\gamma}$ considerably
greater than about 1.2 times the VMD expectation could be considered
as a signal for the presence of large resummation corrections in the
quarkonic anomalous dimension. A measurement of $F_2^{\gamma}$ in the
small-$x$ region will, however, presumably only be possible with the
$e\gamma $ mode of a future $e^+ e^-$ linear collider$\,$%
\footnote{The possible kinematic coverage and necessary detector
 requirements have been studied in ref.~\cite{MV}.}.
Another theoretically cleaner, but experimentally also very difficult
probe would be the structure of highly virtual photons, where the
non-perturbative VMD part is suppressed and the calculable part becomes
more important, cf.~refs.~\cite{BHS}.

\section{Conclusions}
\label{sect7}
\setcounter{equation}{0}

\noindent
The effects of the resummation of the L$x$ and the known NL$x$
small-$x$ contributions to the  flavor-singlet anomalous dimensions
and coefficient functions have been investigated, in a framework
based on the renormalization group equations, for the DIS structure
functions $F_2^{\, p}$ and $F_L^{\, p}$ as well as for the photon
structure function $F_2^{\,\gamma}$. In this approach direct comparisons
are possible with studies of the scaling violations of these structure
functions based on LO and NLO fixed-order calculations. In order to
allow for the most flexible comparison of different approximations to
the all-order evolution equations, their general analytic moment-space
solution has been derived.

The largest small-$x$ corrections to the quark densities and $F_2$ are
due to the resummed quarkonic NL$x$-corrections~\cite{CH}, whereas the
effect of the gluonic terms is marginal here.
For the gluon density, the L$x$ corrections~\cite{LIPAT} are moderately
positive. Both the quarkonic \cite{CC1} and the energy-scale independent
gluonic \cite{CC2} parts of the NL$x$ gluon-gluon anomalous dimension,
on the other hand, cause negative corrections which are that large that
they overcompensate the L$x$-terms. In fact, the latter terms lead to
negative values for the total splitting function $xP_{gg}(x,\alpha_s)$
for $\alpha_s = 0.2$ and $x < 10^{-2}$. This behavior probably signals
the presence of other large positive contributions, either due to the
energy-scale dependent NL$x$-terms or originating in terms of NNL$x$ or
even higher order.

Contributions of NNL$x$ order exhibiting a similar behavior can as
well exist in the case of the quarkonic anomalous dimensions and the
coefficient functions. This is suggested, for example, by the expansion
of the fixed-order anomalous dimensions in powers of $1/N$ which leads
to a good approximation only after three to four terms. Different
ans\"atze for potential less singular terms have been studied
numerically, showing that even the exceedingly large corrections due to
the quarkonic NL$x$-corrections can easily be removed again.

The longitudinal structure function $F_L$ is in addition affected by
the small-$x$ contributions to the coefficient functions $C_L$
\cite{CH}. For lower values of $Q^2$ the corrections become that large
that the positivity constraint $F_L < F_2$ can be violated for
conventional input parton distributions. However, also this resummed
coefficient function is very sensitive to subleading corrections.

All these aspects show that also the next less singular terms need to
be calculated, despite the enormous work that has been carried out so
far to derive the resummed anomalous dimensions and coefficient
functions~[4,8--10,48,49], before firm conclusions on the small-$x$
evolution of singlet structure functions can be drawn. Since
contributions which are even less singular than these ones may, even
then cause relevant corrections, it appears indispensable to compare
the corresponding results to those of future complete fixed-order
calculations. There the  medium and large $x$ terms are fully
contained up to the respective order in $\alpha_s$. If extended to
higher orders, in fact, also the RGE-improved fixed-order perturbation 
theory still seems to remain a viable candidate for the theoretical 
framework in the HERA regime.

The small-$x$ evolution of the real photon's parton structure has been
analyzed in the DIS scheme. It has been shown that this scheme can
be defined, without loss of generality, in such a manner that the
photon-parton splitting functions do not receive any higher-order
resummation corrections at NL$x$ accuracy. Nevertheless the photon
structure function $F_2^{\,\gamma}$ can provide an additional
laboratory for studying the possible effect of small-$x$ resummations,
as the characteristic, calculable inhomogeneous solution of the
evolution equations probes the low-$x$ hadronic anomalous dimensions
in a unique way: it comes closer to a local probe of the small-$x$
quarkonic splitting functions than any inclusive hadronic quantity.
Unfortunately, this particularly interesting contribution is likely
to be dominated by the hadron-like vector-meson-dominance part which
behaves completely analogous to the photon structure and hence
introduces the same uncertainties and limitations due to the interplay
of the anomalous dimensions and the non-perturbative initial 
distributions.

\vspace{5mm}
\noindent
{\bf Acknowledgements :} Our thanks are due to P. S\"oding for his
constant support of this project. We would also like to thank
M.~Ciafaloni, W.~van Neerven, D.~Robaschik, G.~Camici, and S.~Riemersma
for useful discussions. This work was supported in part by the German
Federal Ministry for Research and Technology (BMBF) under contract
\mbox{No.\ 05 7WZ91P (0).}


\newpage
\section*{Figure captions}
\label{figcap}
\begin{description}
\item[{\bf Fig.~1}] The real and imaginary parts of the perturbative
 branch of $\gamma_L$ \cite{LIPAT} as a function of $\rho = N/\ab$.
 The the dash-dotted lines are the contours through the singularities,
 Eq.~(\ref{eqBRA}).
\item[{\bf Fig.~2}] {\bf (a)} The cumulative effect of the available
 contributions on the splitting function $xP_{qg}(x)$ for $\alpha_s =
 0.2$. The fixed-order results are supplemented by the small-$x$
 resummed NL$x$ corrections \cite{CH} beyond NLO. Also shown are the
 modifications induced by the subleading terms (C) and (D) of Sect.~4.
 {\bf (b)} As in (a) but for the convolution of $P_{qg}$ with a typical
 shape of an hadronic gluon density.
\item[{\bf Fig.~3}] {\bf (a)} The cumulative effect of the available
 terms for splitting function $xP_{gg}(x)$ for $\alpha_s = 0.2$. The
 fixed-order results are successively supplemented by the higher-order
 small-$x$ resummed L$x$ correction \cite{LIPAT}, the $q\overline{q}$
 contribution to the NL$x$ term \cite{CC1}, and the gluonic NL$x$
 energy-scale independent terms \cite{CC2}. {\bf (b)} As (a) but for
 the convolution of $P_{gg}$ with a shape of an hadronic gluon density.
\item[{\bf Fig.~4}] Integration contour in the complex $N$-plane for
 the inverse Mellin transformation (\ref{minv}) relative to the
 locations of the singularities of typical initial parton distributions
 (full circles), and those of the fixed-order (crosses) and resummed
 anomalous dimensions (open and closed diamonds for different values
 of $Q^2$).
\item[{\bf Fig.~5}]  The small-$x$ evolution of the proton's flavor-%
 singlet quark and gluon distributions in LO and NLO perturbative QCD.
 Two approaches to the solution of the evolution equations are compared
 in the NLO case, cf.~Sect.~5. Also shown is the result of a small-$x$
 approximation (NL$x_q$, see the text) of the LO splitting functions.
 The behavior of non-singlet quantities is illustrated by the total
 valence quark distribution $x(u_v\! +\! d_v)$.
\item[{\bf Fig.~6}]  The effects of the successive small-$x$
 approximations (\ref{eqAPP}) to the LO anomalous dimensions
 $\gamma_0$ (left) and to the NLO corrections $\gamma_1$ (right) on the
 low-$x$ evolution of $x\Sigma^p$ and $xg^p$, cf.~Eqs.~(4.1) and
 (4.2). All results are displayed at $Q^2 = 100 \mbox{ GeV}^2$ relative
 to the respective full calculations presented in Fig.~5.
\item[{\bf Fig.~7}]  The resummed small-$x$ evolution of the singlet
 quark and gluon densities as compared to the NLO results. The L$x$
 \cite{LIPAT} and NL$x$ [8--10]           contributions are
 successively included, with the momentum sum rule implemented via
 prescription (A) of Sect.~4. The results for $Q^2 = 10$ and $100
 \mbox{ GeV}^2$ have been multiplied by the factors indicated in the
 plots.
\item[{\bf Fig.~8}]  {\bf (a)} The impact of the resummed gluonic
 splitting functions $P_{gg}$ and $P_{gq}$ of refs.~\cite{LIPAT,CH,CC1}
 on $x\Sigma^p$ and $xg^p$ at $Q^2 = 100 \mbox{ GeV}^2$. The NL$x_q$
 results of Fig.~7 have been chosen as reference.  {\bf (b)} The
 offsets between the iterated and the truncated solutions of the
 evolution equations, see Sect.~5, for the NLO case and various
 resummation approximations. In all other figures the truncated
 solutions have been employed.
\item[{\bf Fig.~9}]  The possible effects of subleading corrections
 to the resummed anomalous dimensions, exemplified by the momentum
 sum-rule prescriptions (C) and (D) of Eq.~(\ref{eq63}), on the
 \mbox{small-$x$} evolution of the proton's parton densities. The
 results using the $\delta (1-x)$ subtractions (A) are as in Fig.~7.
 The MRS(A$'$) \cite{MRSA} initial distributions (transformed to the
 DIS scheme) have been employed as in all other proton figures.
\item[{\bf Fig.~10}]  The small-$x$ behavior of the resummed structure
 functions $F_2$ and $F_L$ for the parton evolutions shown in the
 previous figure in comparison with the NLO results. The upper $F_L$
 curves include the resummed coefficient functions $C_L$ \cite{CH},
 cf.\ Eq.~(\ref{eq64}), the lower ones illustrate the possible impact
 of a subleading contribution to $C_L$.
\item[{\bf Fig.~11}] The small-$x$ evolution of the photon's singlet
 quark and gluon distributions in leading and next-to-leading order,
 starting from the NLO parametrization of \cite{GRVph2} at $Q_0^2 =
 1 \mbox{ GeV}^2$ as in all following figures. The hadronic
 (vector-meson-dominance induced) components are compared to the full
 results at NLO.
\item[{\bf Fig.~12}] The resummed small-$x$ evolution of $x\Sigma
 ^{\gamma}$ and $xg^{\,\gamma}$ to $Q^2 = 100 \mbox { GeV}^2$ as
 compared to the NLO results. The L$x$ \cite{LIPAT} and NL$x$
 [8--10]           contributions are successively included, with
 the momentum sum rule implemented via prescription (A) of Sect.~4.
 The effects on the photon-specific inhomogeneous solution are
 displayed separately.
\item[{\bf Fig.~13}] The possible effects of subleading corrections
 to the resummed evolution kernels, exemplified by the momentum
 sum-rule prescriptions (C) and (D) of Eq.~(\ref{eq63}), on the
 \mbox{small-$x$} evolution of the photon structure function $F_2^{\,
 \gamma}$ and the photon's gluon distribution.
\end{description}
\newpage
\vfill
\vspace*{0.6cm}
\centerline{\epsfig{file=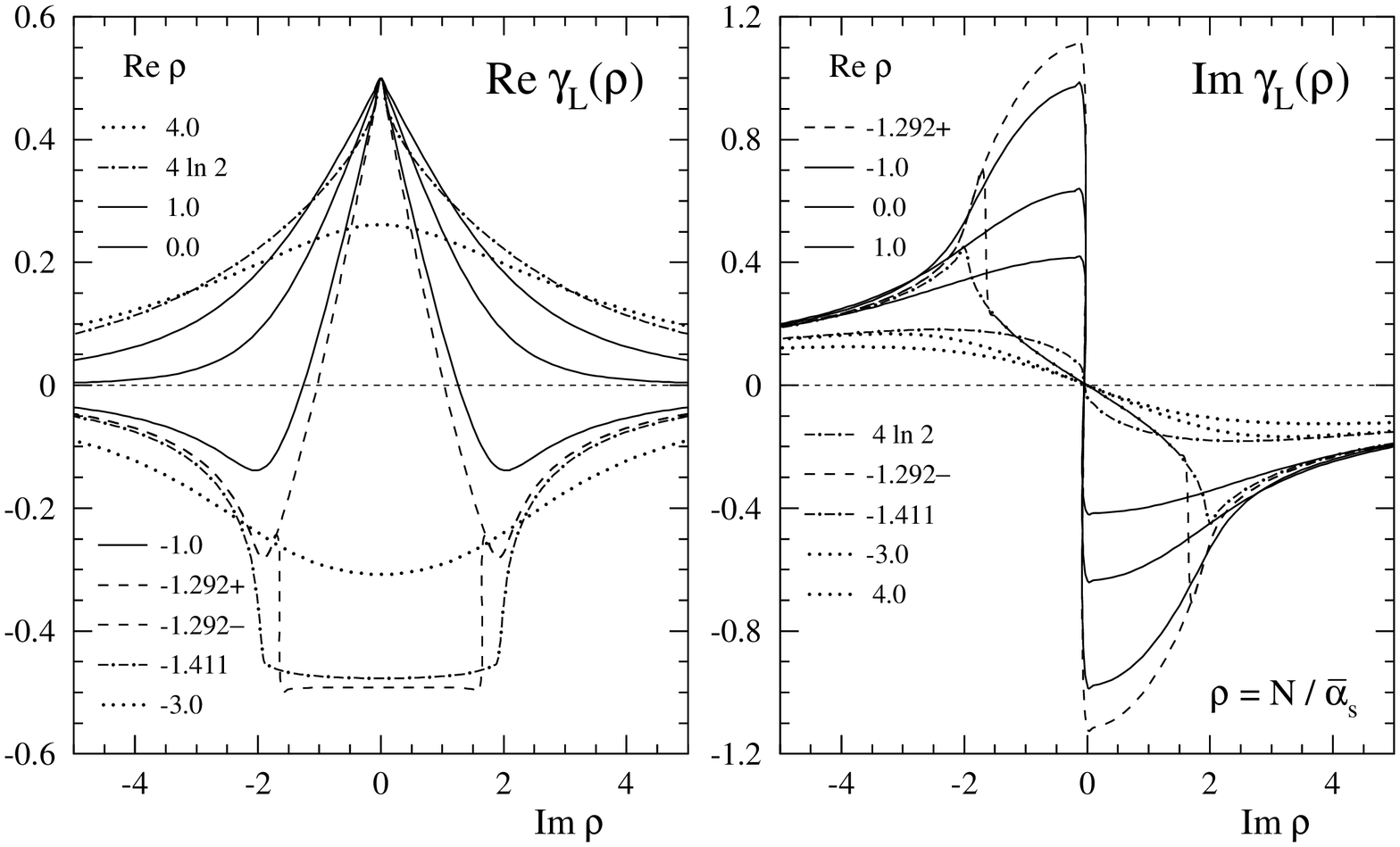,width=21.5cm,angle=90}}
\vspace{0.4cm}
\centerline{\bf Fig.~1}
\vfill

\newpage
\vfill
\vspace*{0.6cm}
\centerline{\epsfig{file=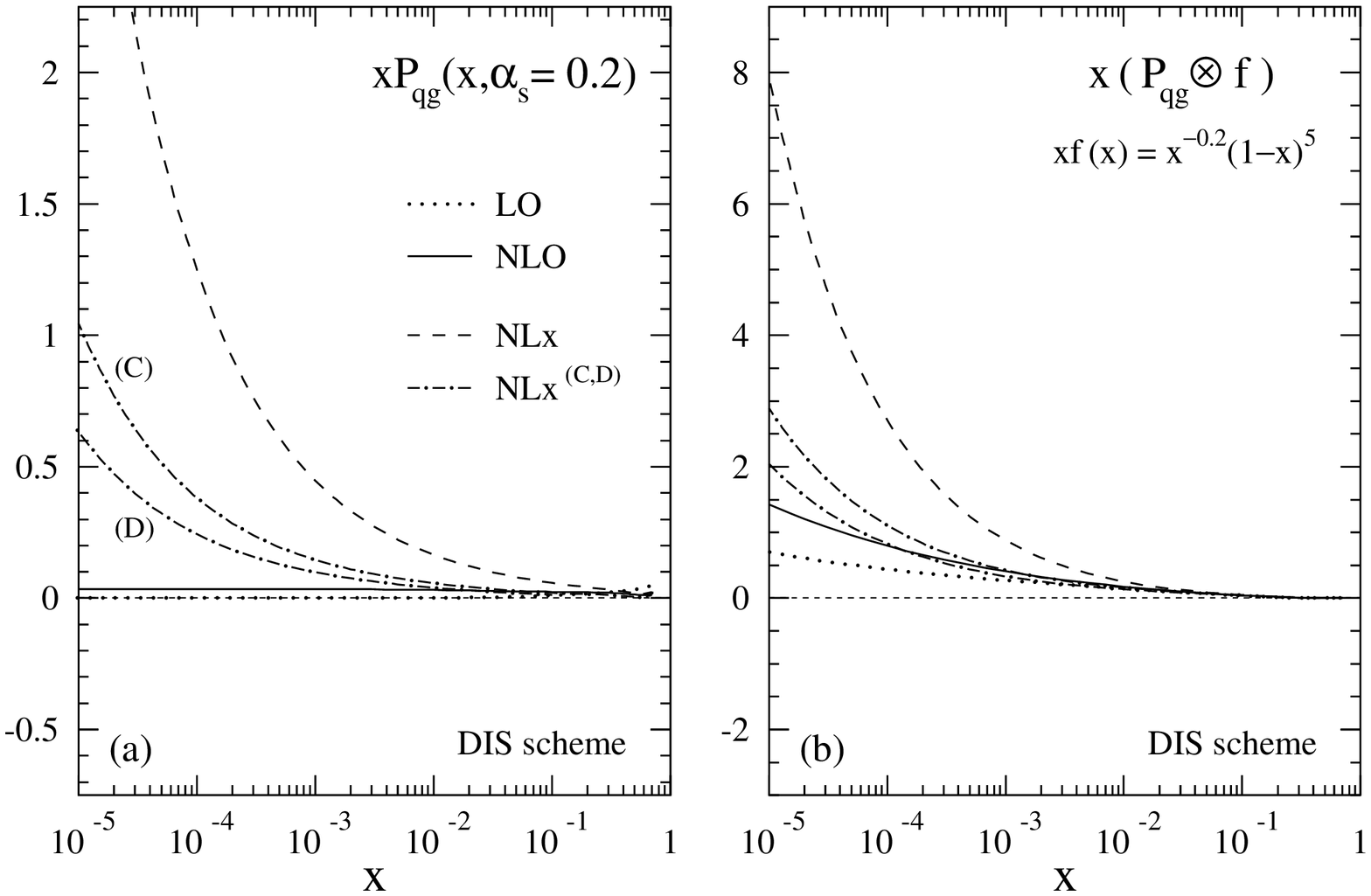,width=21.5cm,angle=90}}
\vspace{0.4cm}
\centerline{\bf Fig.~2}
\vfill

\newpage
\vfill
\vspace*{0.6cm}
\centerline{\epsfig{file=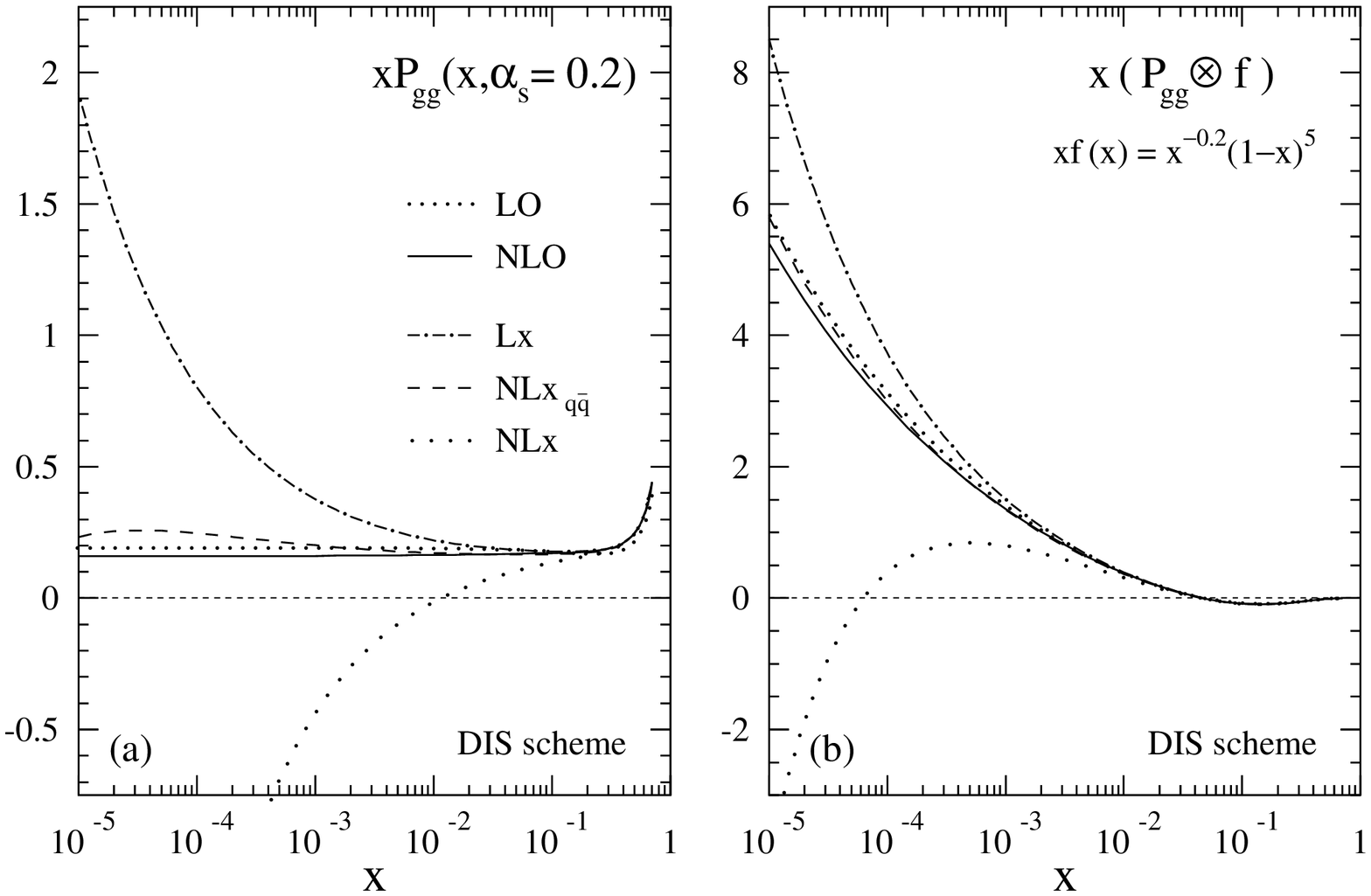,width=21.5cm,angle=90}}
\vspace{0.4cm}
\centerline{\bf Fig.~3}
\vfill

\newpage
\vfill
\vspace*{5.0cm}
\centerline{\epsfig{file=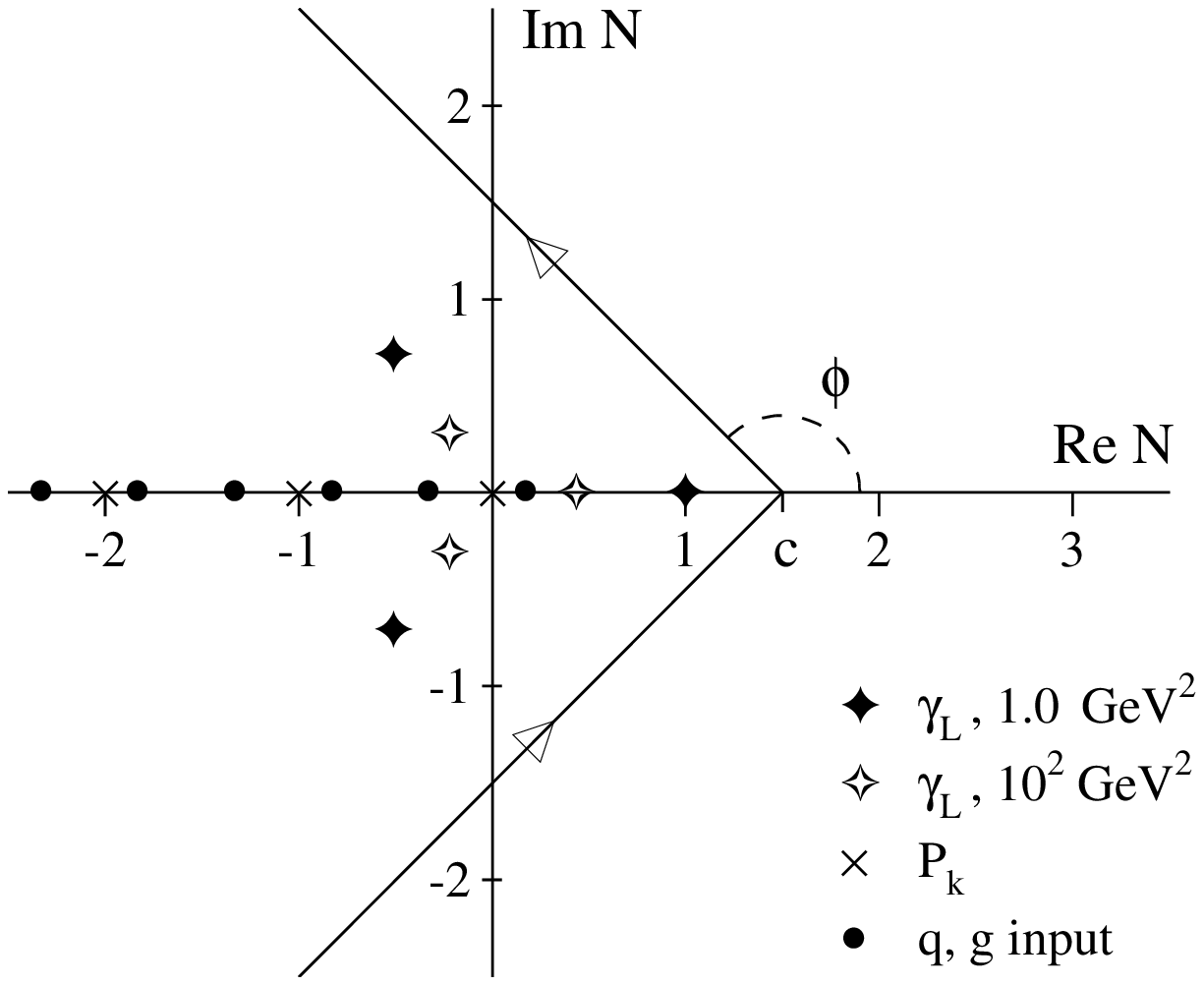,width=14cm,angle=0}}
\vspace{0.4cm}
\centerline{\bf Fig.~4}
\vfill

\newpage
\vfill
\vspace*{0.6cm}
\centerline{\epsfig{file=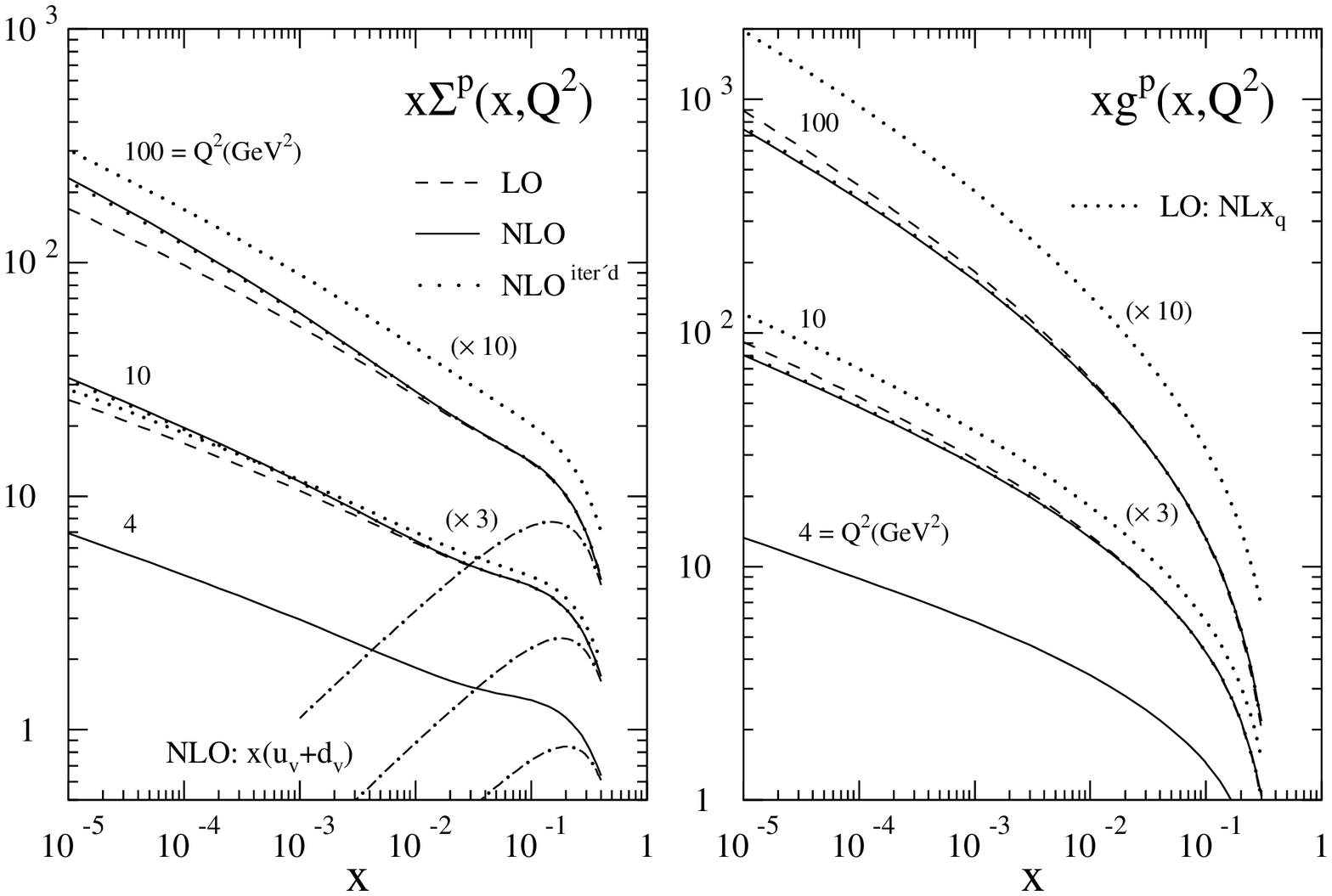,width=21.5cm,angle=90}}
\vspace{0.4cm}
\centerline{\bf Fig.~5}
\vfill

\newpage
\vfill
\vspace*{0.6cm}
\centerline{\epsfig{file=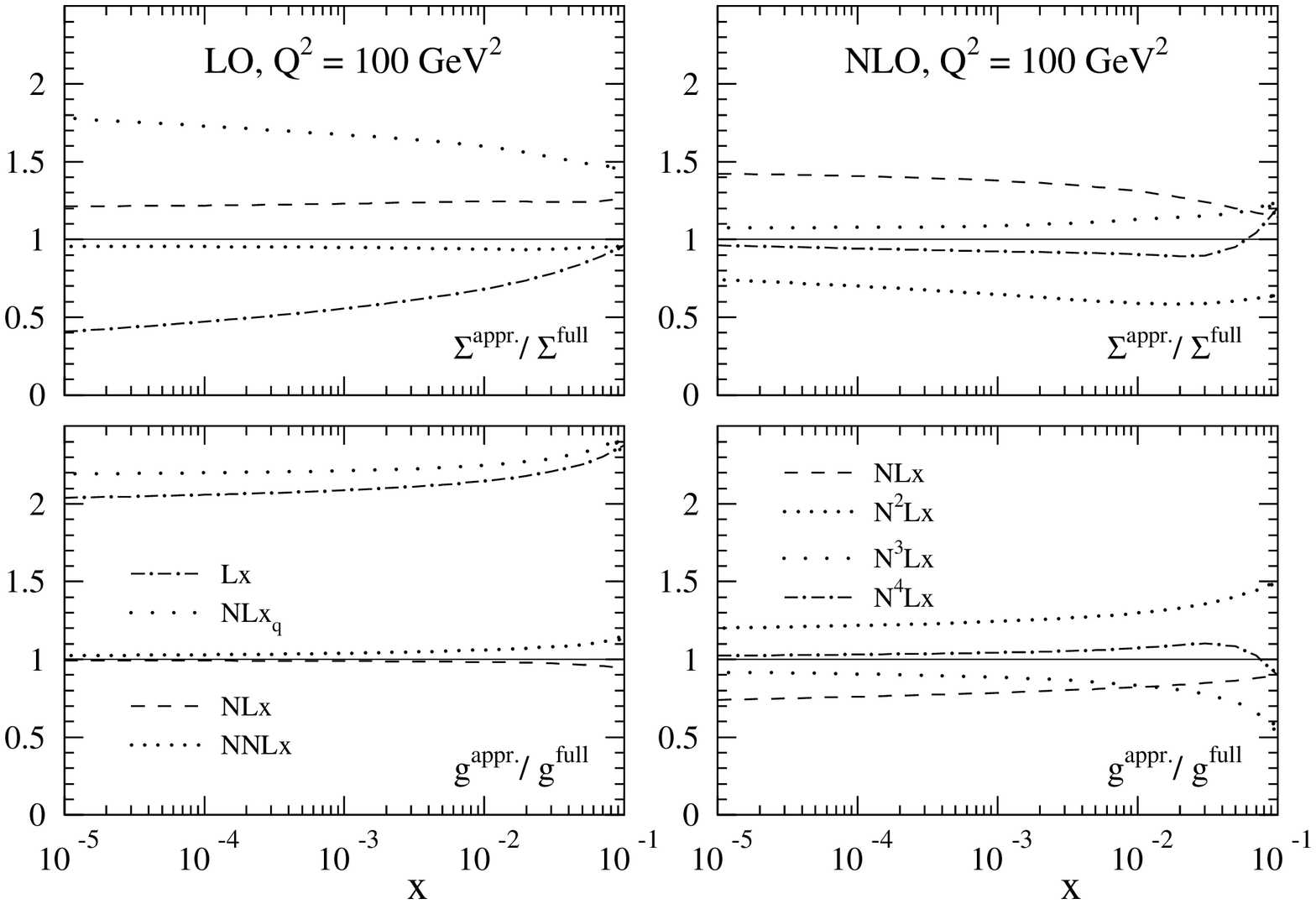,width=21.5cm,angle=90}}
\vspace{0.4cm}
\centerline{\bf Fig.~6}
\vfill

\newpage
\vfill
\vspace*{0.6cm}
\centerline{\epsfig{file=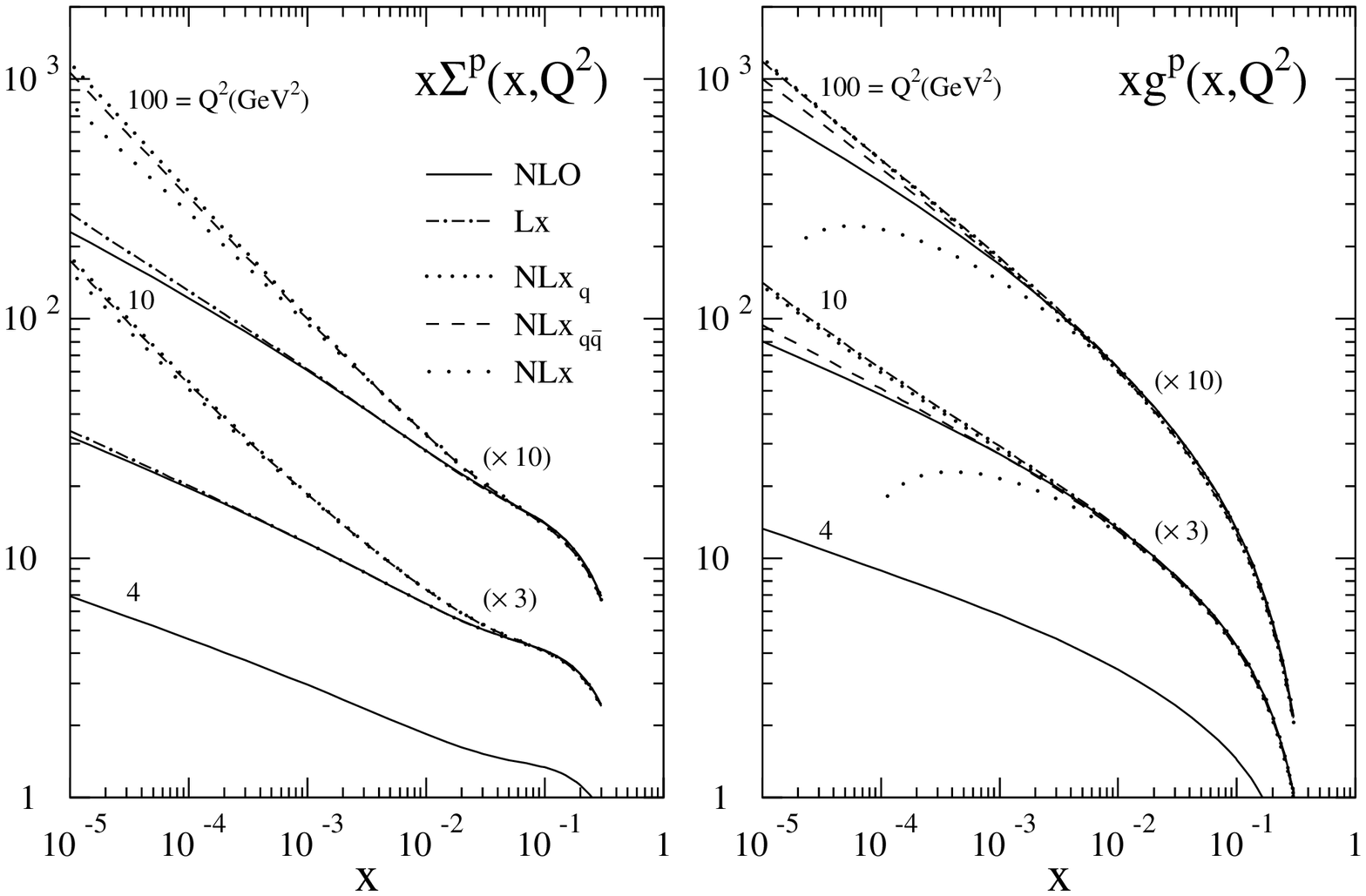,width=21.5cm,angle=90}}
\vspace{0.4cm}
\centerline{\bf Fig.~7}
\vfill

\newpage
\vfill
\vspace*{0.6cm}
\centerline{\epsfig{file=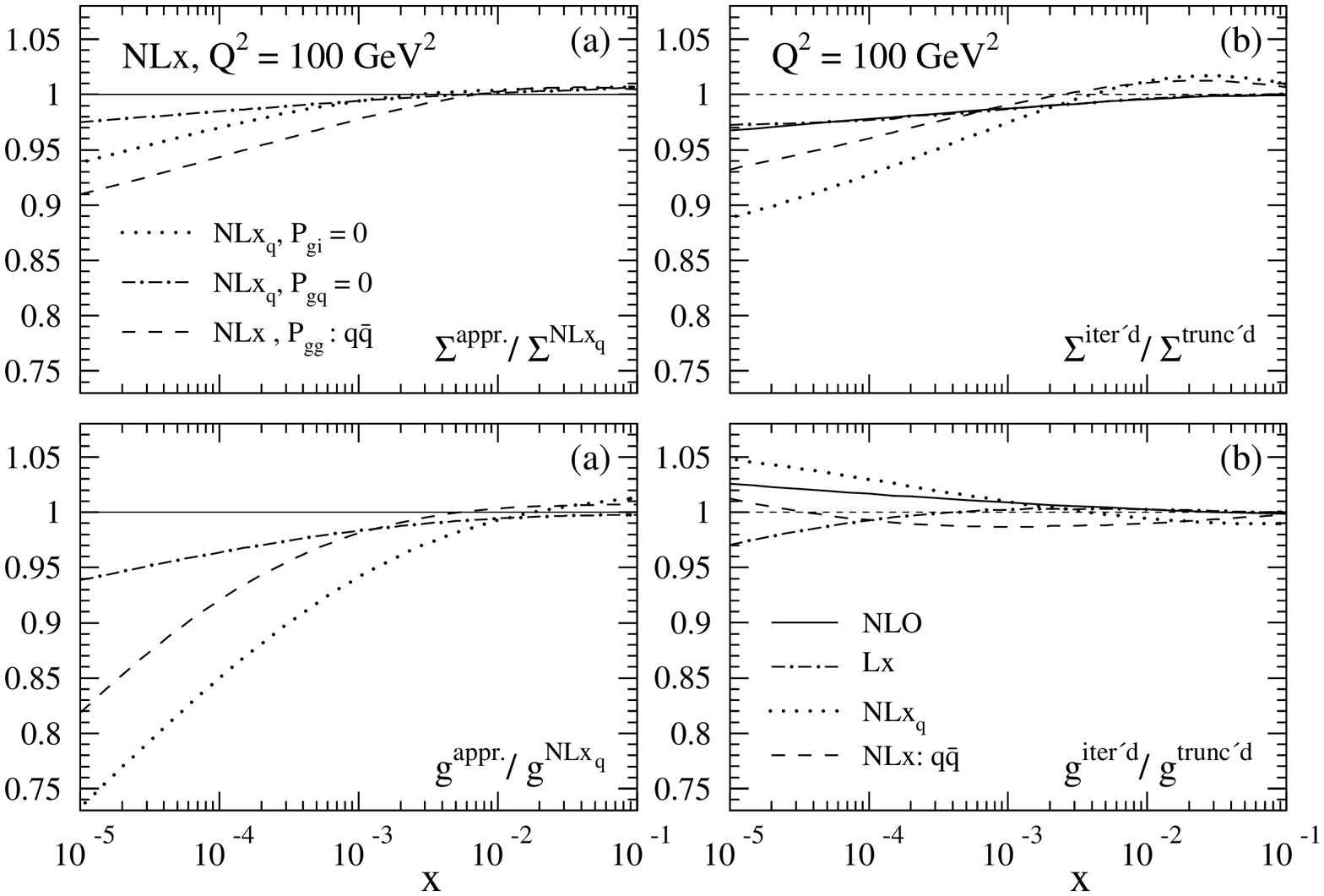,width=21.5cm,angle=90}}
\vspace{0.4cm}
\centerline{\bf Fig.~8}
\vfill

\newpage
\vfill
\vspace*{0.6cm}
\centerline{\epsfig{file=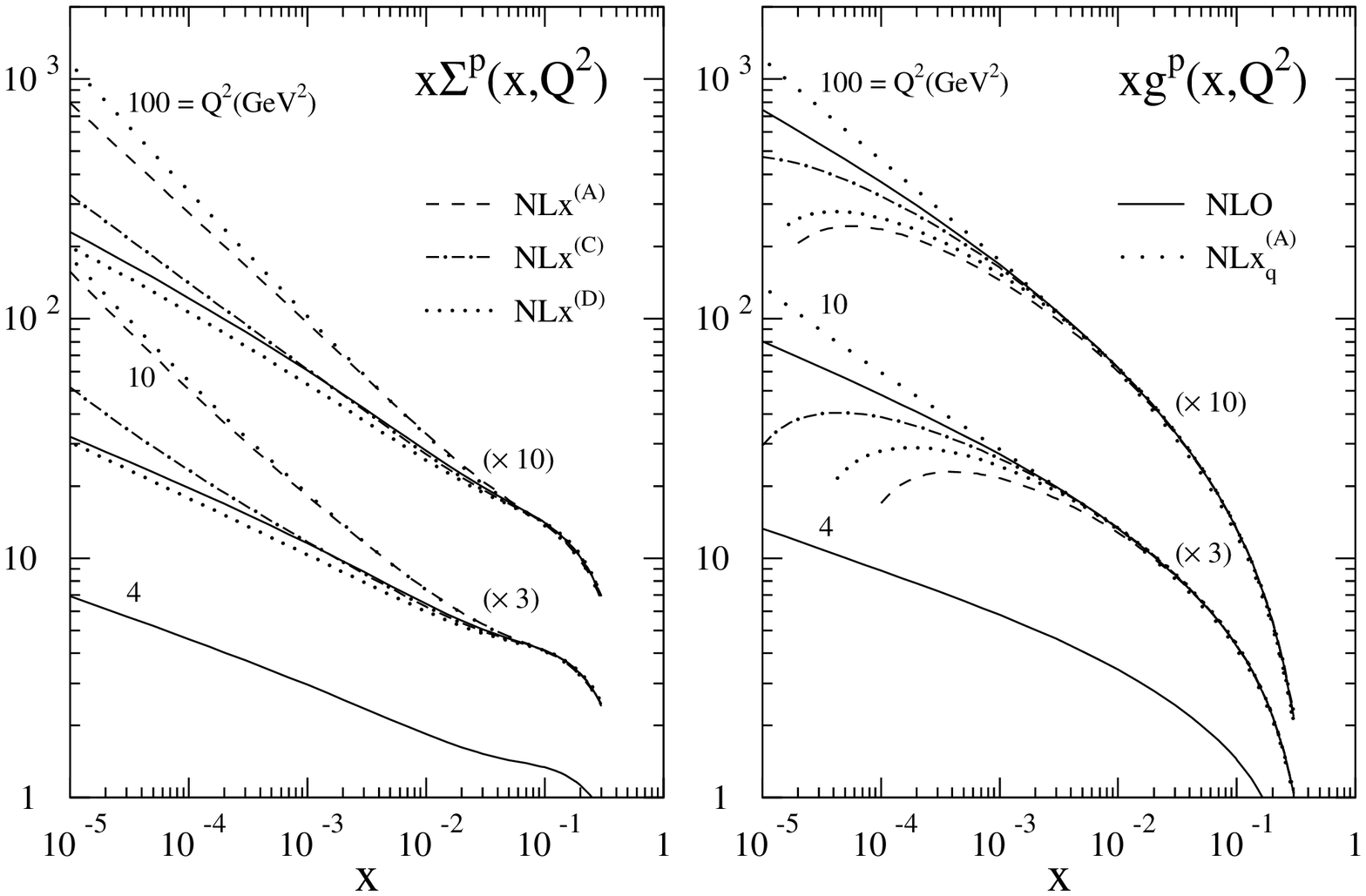,width=21.5cm,angle=90}}
\vspace{0.4cm}
\centerline{\bf Fig.~9}
\vfill

\newpage
\vfill
\vspace*{0.6cm}
\centerline{\epsfig{file=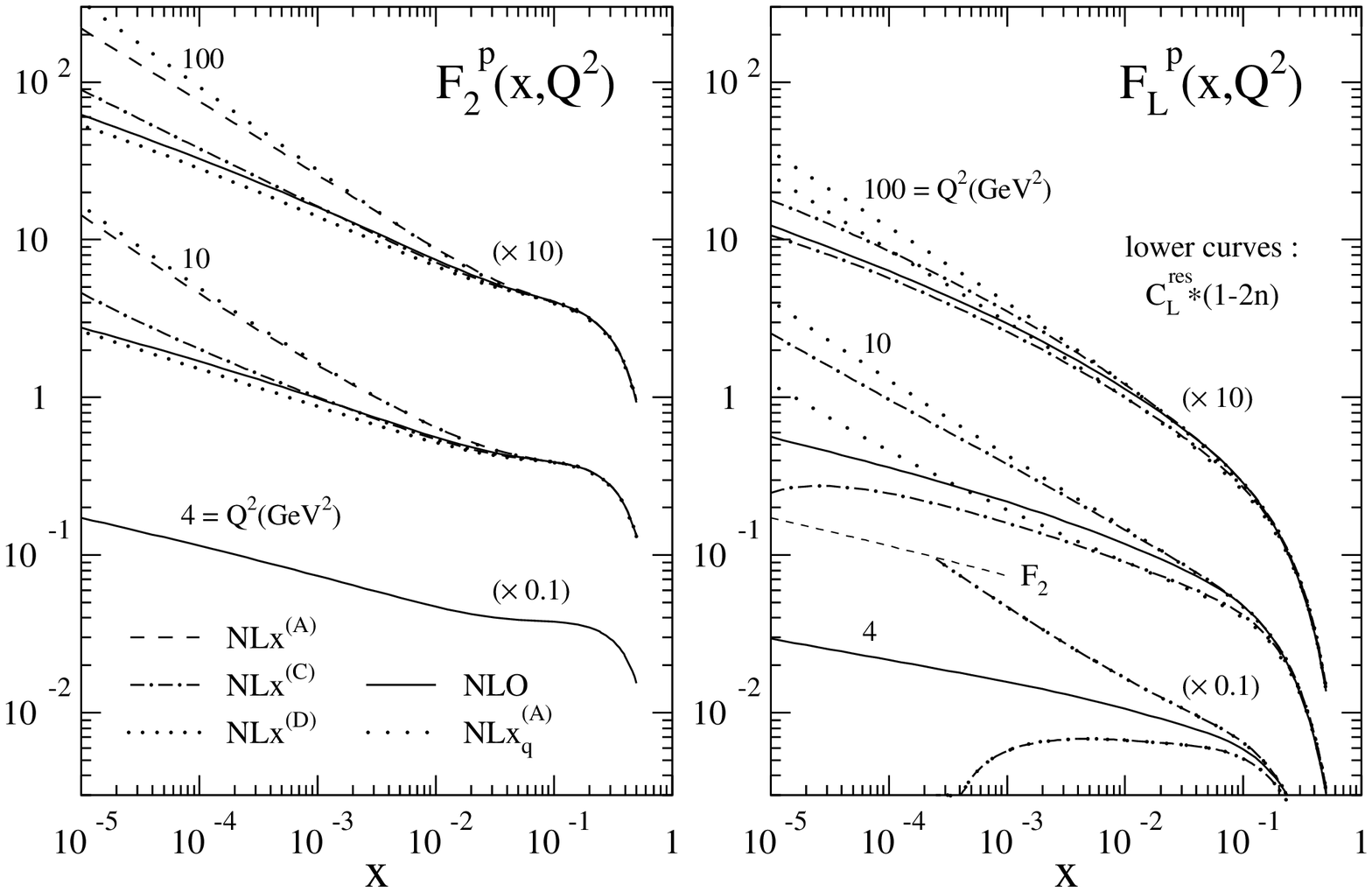,width=21.5cm,angle=90}}
\vspace{0.4cm}
\centerline{\bf Fig.~10}
\vfill

\newpage
\vfill
\vspace*{0.6cm}
\centerline{\epsfig{file=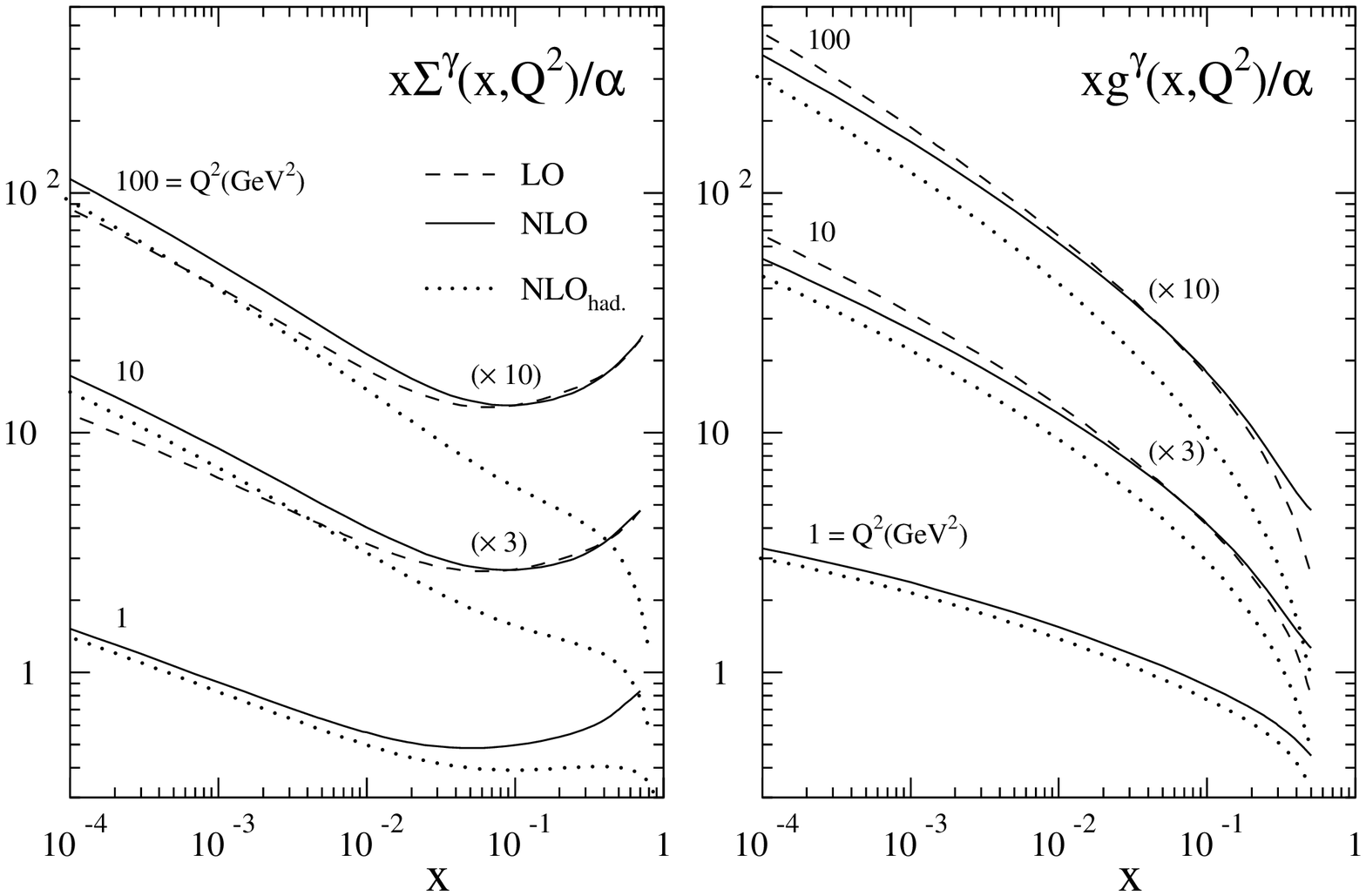,width=21.5cm,angle=90}}
\vspace{0.4cm}
\centerline{\bf Fig.~11}
\vfill

\newpage
\vfill
\vspace*{0.6cm}
\centerline{\epsfig{file=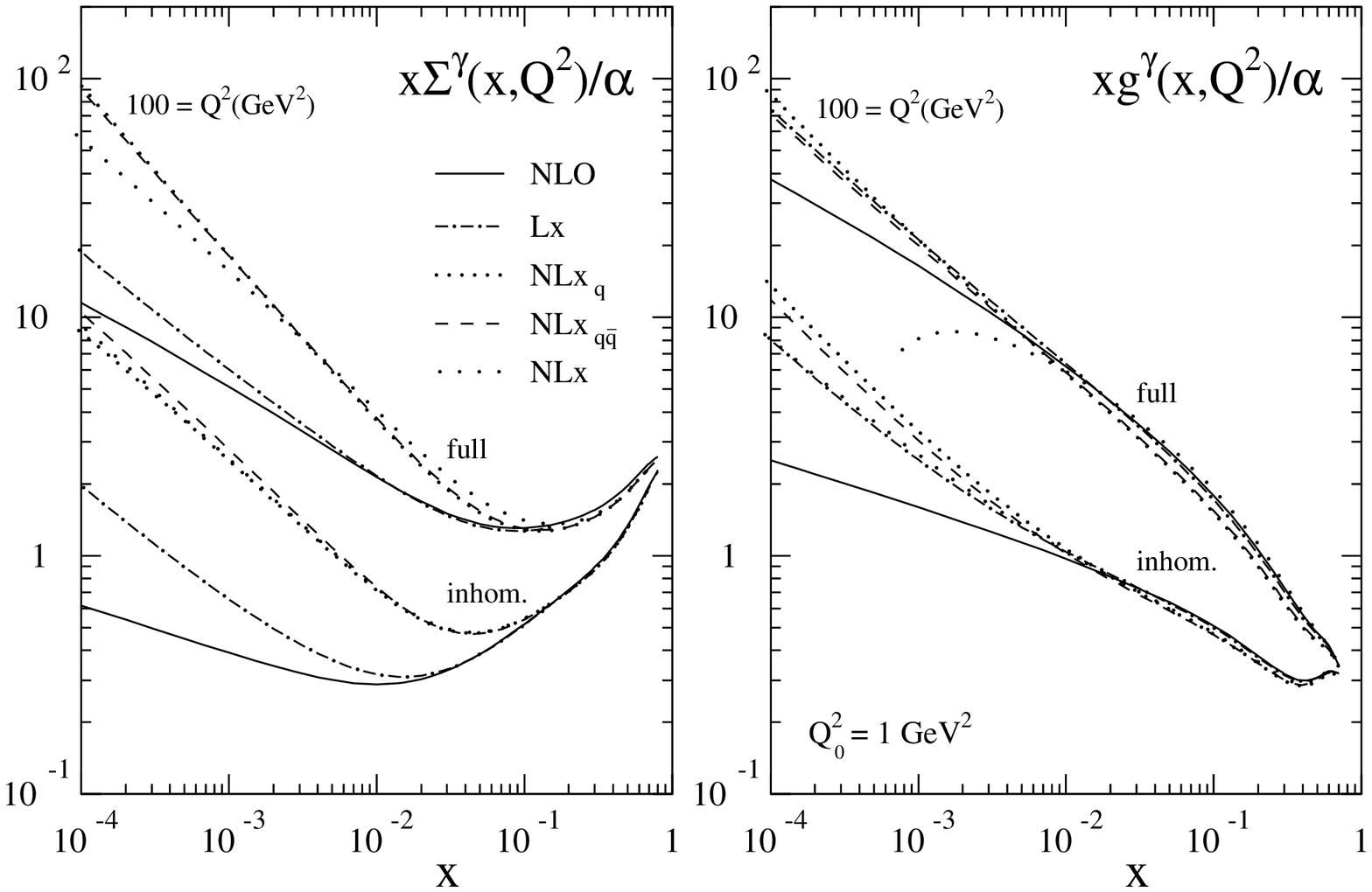,width=21.5cm,angle=90}}
\vspace{0.4cm}
\centerline{\bf Fig.~12}
\vfill

\newpage
\vfill
\vspace*{0.6cm}
\centerline{\epsfig{file=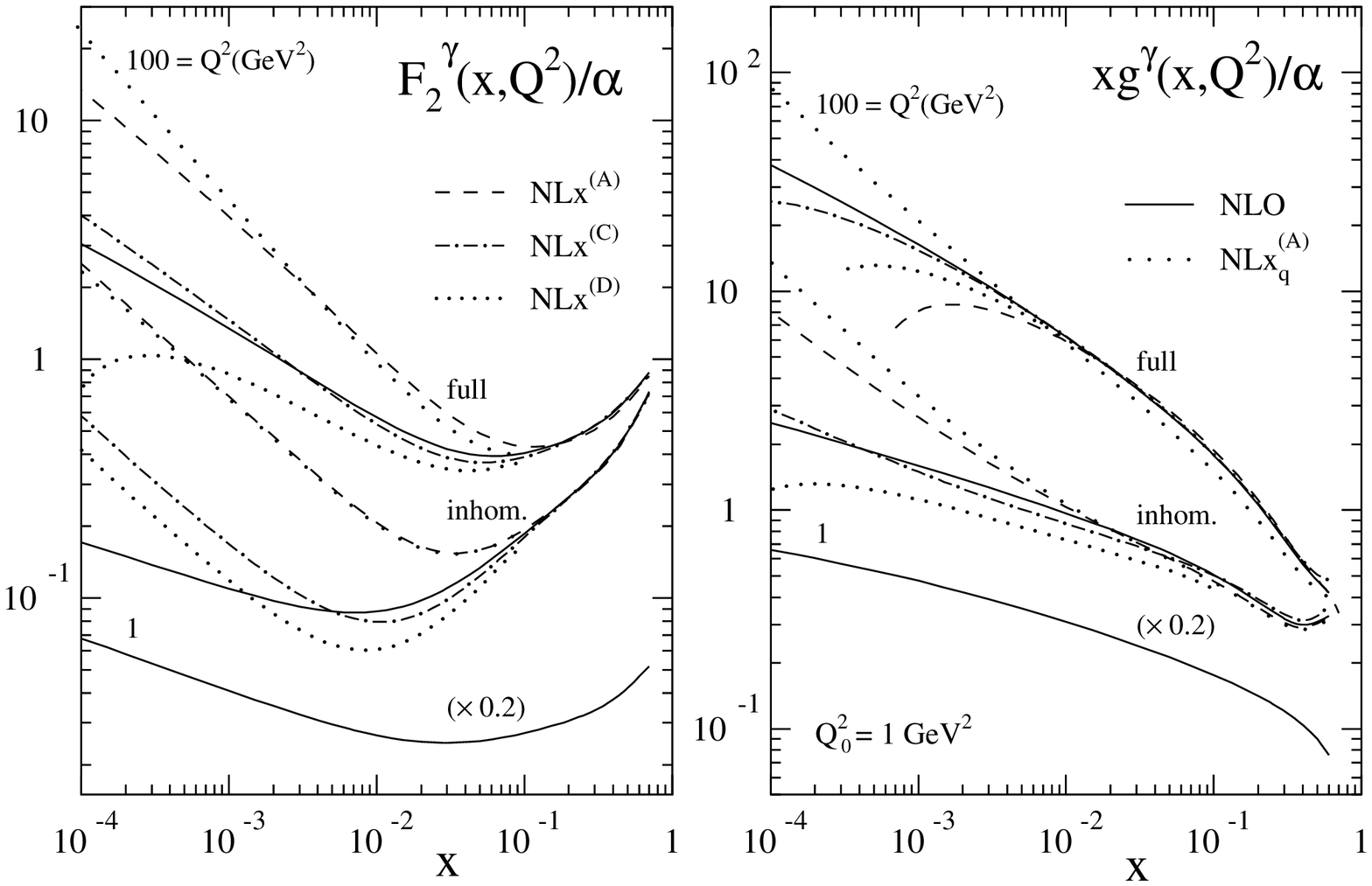,width=21.5cm,angle=90}}
\vspace{0.4cm}
\centerline{\bf Fig.~13}
\vfill

\end{document}